\documentclass{article}
\usepackage{authblk}
\usepackage[tbtags]{amsmath}
\usepackage{bm}
\usepackage[onehalfspacing]{setspace}
\providecommand{\keywords}[1]{\textbf{\textit{Index terms---}} #1}
\usepackage{textcomp,gensymb}
\usepackage{quoting}
\usepackage{epsfig}
\usepackage{color}
\usepackage{amsmath}
\usepackage{cite}
\usepackage{graphicx}
\usepackage{subfig}

\begin{document}

\title{\textbf{Dynamic~Beam~Shaping~Using~a~Dual-Band~Metasurface-Inspired~Electronically~Tunable~Reflectarray~Antenna}}

\author{Amin~Tayebi%
\thanks{E-mail: \texttt{tayebiam@egr.msu.edu}; Corresponding author}}
\affil{\textit{Department of Electrical and Computer Engineering, and Department of Physics and Astronomy, Michigan State University, East Lansing, MI, 48824 USA}}

\author{J.~Tang, P.~Paladhi, L.~Udpa, S.~Udpa, and E.~J.~Rothwell}
\affil{\textit{Department of Electrical and Computer Engineering, Michigan State University, East Lansing, MI, 48824 USA}}

\date{Dated: \today}

\maketitle

(A shorter version of this paper is published in IEEE Transactions on Antennas and Propagation. DOI: 10.1109/TAP.2015.2456939)

\begin{abstract}
An electronically reconfigurable dual-band-reflectarray antenna is presented in this paper.
The tunable unit cell, a ring loaded square patch with a single varactor diode connected across the gap between the ring and the patch, is modeled using both a full-wave solver and an equivalent circuit. The parameters of the equivalent circuit are calculated independently of the simulation and experiment using analysis techniques employed in frequency selective surfaces. The reflection phase of the proposed unit cell is shown to provide an excellent phase range of 335$^{\circ}$ in F band and 340$^{\circ}$ in S  band. Results from the analysis are used to design and build a 10x10 element reflectarray antenna. The high tuning phase range of each element allows the fabricated reflectarray to demonstrate a very broad steering range of up to $\pm$60$^{\circ}$ in both frequency bands.
\end{abstract}


\section{Introduction}

The reflectarray antenna, first introduced in 1963, can be thought of as a reflecting surface where the elements on the array determine the reflection properties of the surface \cite{Berry}. That is, the amplitude and phase change between the incident and the reflected fields, at any point on the surface of the reflectarray, are determined by the elements of the array. Similar to antenna arrays, in order to achieve the desired radiation pattern, the proper phase shift has to be assigned to each element of the array, which can be done by appropriately choosing the parameters of individual elements. An array of variable-length, open-ended waveguides was used in \cite{Berry} to demonstrate the reflectarray principle.

Later, with the advent of microstrip technology and the extension to microstrip arrays, reflectarrays became easier to build. The behavior of a microstrip array is defined by its printed resonating elements. First generation microstrip reflectarrays comprised patch elements with variable-length transmission lines attached to the patches where the reflection from the open end of the line determined the phase shift of the element \cite{Javor}. Rectangular patches of variable lengths are used in \cite{Pozar} since the reflection phase changes with the length of the patch. A variety of microstrip reflectarrays have been developed using other radiating elements, such as variable-length crossed dipoles \cite{Kelkar} and rotated patches \cite{Huang}, that lead to circularly-polarized reflectarrays.

Research on reflectarrays continued to advance with the introduction of dual-band reflectarrays, in which each unit cell comprises two different elements, each accountable for radiating in one of the bands. Crossed-dipoles of variable sizes printed on a single layer \cite{Kelkar}, patches with variable sizes on two stacked layers \cite{Encinar} and double ring elements \cite{Huang3} are all examples of dual-band reflectarray antennas.

An extra dimension of versatility is added to this group of antennas by their ability to dynamically adjust element scattering properties in order to alter the radiation pattern, and thereby produce tunable reflectarrays. Reconfigurable elements are achieved via several techniques. In \cite{Inam} a pin diode is incorporated in the unit cell; the ON and OFF states of the diode alter the electrical length of the element and hence provide tunablity. Beam shaping is achieved by means of RF MEMS in \cite{Bayraktar}. Technologies such as liquid crystals \cite{Hu}, ferroelectric materials \cite{Romanofsky} and graphene \cite{Carrasco} have been used as well. A recent review on the advantages and disadvantages of different technologies can be found in \cite{Hum1}. These techniques can be combined with dual-band unit cells to produce tunable dual-band reflectarrays. In \cite{Guclu} and \cite{Moghadas} two tunable dual-band unit cells based on RF MEMS are proposed as reflectarray antennas, though the array has not yet been built.

Beam steering using high impedance surfaces (HIS) is a relatively new and fascinating area in the field of reflectarrays. A varactor tunable HIS element is presented in \cite{Mias}. Despite their promising performance, the size of the periodic element is often much smaller than the operating wavelength, therefore more elements are required on the surface of the array. This leads to an increase in the cost and complexity of the array.

In this paper varactor diodes are used to create tunable array elements, since they can be easily incorporated into the unit cell and are more cost-effective, and more reliable, than the previously mentioned techniques. Recent developments in unit cells that exploit varactor diodes for phase adjustability have shown promising results \cite{Hum2}, \cite{Rodrigo}. The unit cell in \cite{Hum2}, which uses two varactor diodes, is capable of providing a phase shift of up to 320$^{\circ}$. Using two varactors per unit cell is not only costly, but the biasing circuit fabricated on the reflecting surface of the antenna increases loss and lowers the scattering amplitude. In \cite{Rodrigo}, frequency tunability is achieved using a single varactor, but at the cost of a lower phase change of 270$^{\circ}$, and the additional need for two RF switches and a corresponding complexity of fabrication.

In the following sections, a dual-band metamaterial-inspired unit cell is introduced that uses a single varactor diode to dynamically adjust the phase of the reflected electric field in the two frequency bands \cite{Tayebi}, \cite{Tayebi2}. The unit cell is modeled and analyzed, and the results are used to design a reflectarray antenna. Finally, a prototype of the antenna is constructed, and the subsequent measurements are used to verify the predicted properties of the proposed antenna.

\section{The Theory of Reflectarrays}

Several techniques have been used for the analysis of reflectarrays. Numerical methods, such as finite difference time domain (FDTD) and finite element method (FEM) modeling, described in \cite{Cadoret} and \cite{Bardi} respectively, provide accurate results, although they are time consuming. In this paper, similar to \cite{Berry}, a simple yet effective theory based on the concept of impedance surfaces is utilized.

\begin{figure}[!ht]
\begin{center}
\includegraphics[keepaspectratio = true, width = 2.75in, clip = true]{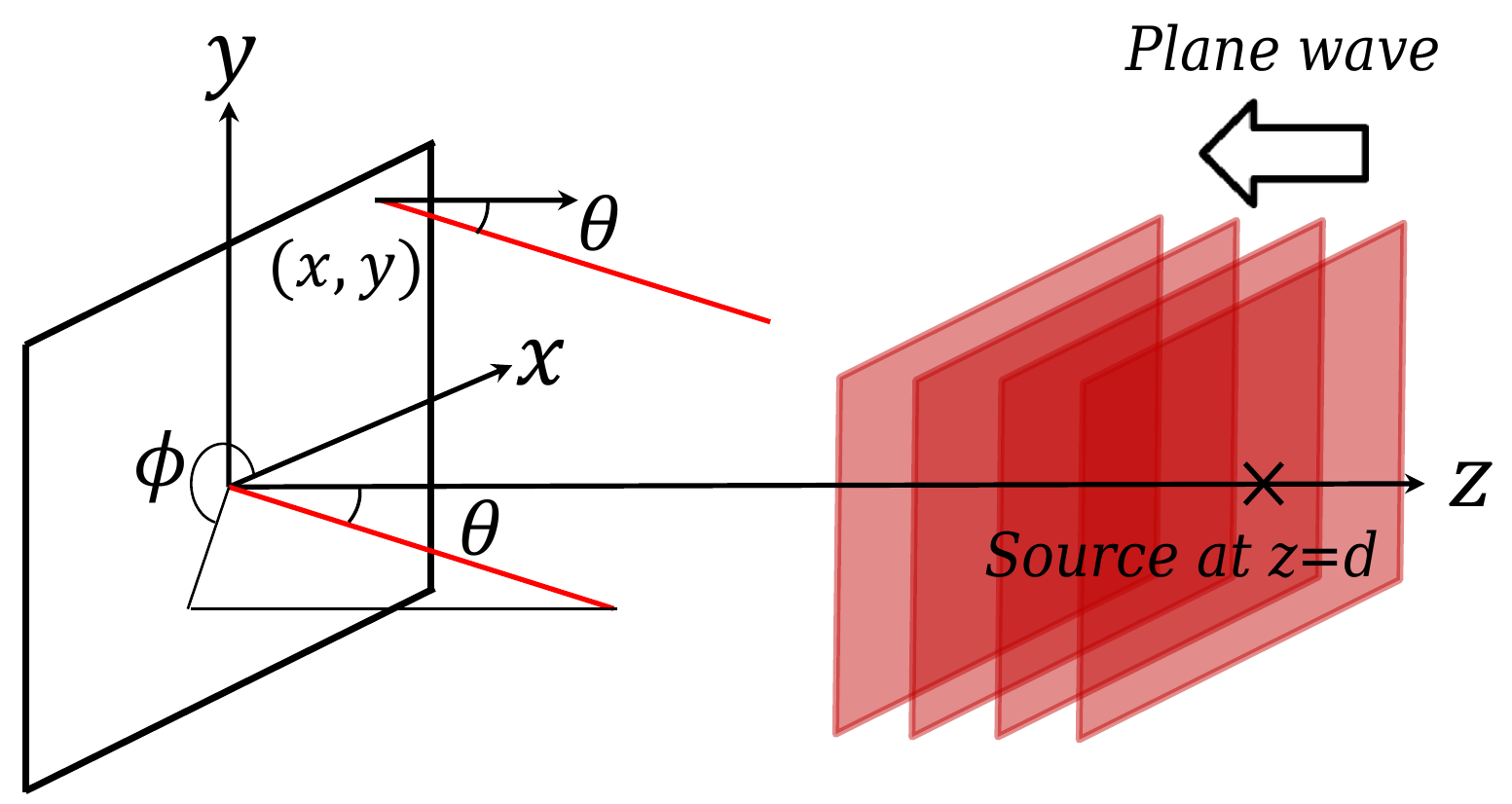}
\caption{Reflectarray geometry.}  \label{Sketch}
\end{center}
\end{figure}

Consider the infinitely large reflectarray shown in Fig. \ref{Sketch}. For simplicity, it is assumed that the incident field is a plane wave normal to the reflectarray and linearly polarized along the $y$-axis. This is a good approximation when the reflectarray feed is far away from the surface of the array or is capable of generating plane waves, such as with lensed horn antennas. Under this assumption the electric field on the surface of the array is expressed as
\begin{equation}\label{eqn1}
\vec{E}_{i} =  E_{0} e^{-jkd} \hat{y},
\end{equation}
where $E_{0}$ is the amplitude of the wave, $k=2\pi/\lambda_{0}$ is the wavenumber corresponding to the free space wavelength $\lambda_{0}$, and $d$ is the distance of the source from the array.

Exploiting the concept of surface impedance \cite{Schelkunoff}, the reflected electric field is
\begin{equation}\label{eqn2}
\vec{E}_{r} = \Gamma(x,y) \vec{E}_{i},
\end{equation}
where $\Gamma$ is the reflection coefficient at $(x,y)$ on the surface of the array as given by
\begin{equation}\label{eqn3}
\Gamma(x,y) = \frac{Z_{s}(x,y)-\eta_{0}}{Z_{s}(x,y)+\eta_{0}} .
\end{equation}
Here $\eta_{0}$ is the free space impedance of plane waves and $Z_{s}$ is the surface impedance at point $(x,y)$. The parameter $Z_{s}$ projects the effects of various phenomena, such as absorption, radiation, and scattering that occur in the region $z\leq0$, and projects them onto the surface $=0$.

Assuming the reflectarray is lossless and passive, the surface impedance is purely reactive such that
\begin{equation}\label{eqn4}
Z_{s}= j X_{s}
\end{equation}
An appropriate expression for $X_{s}$ is derived in section \MakeUppercase{\romannumeral 3}, where the equivalent circuit of the unit cell is introduced. For now, it is sufficient to assume that $X_{s}\geq0$; $X_{s}=0$ occurs in the case of a perfectly reflecting surface (short circuit).

Using (\ref{eqn3}) and (\ref{eqn4}), the reflected electric field on a plane infinitesimally close to the surface of the array is
\begin{equation}\label{eqn5}
\indent \vec{E}_{r} = \frac{jX_{s}(x,y)-\eta_{0}}{jX_{s}(x,y)+\eta_{0}} E_{0}e^{-jkd} \hat{y}.
\end{equation}
Two limiting cases are of practical interest. First, when $X_{s}\ll\eta_{0}$ a surface with properties similar to a perfect electrical conductor (PEC) is produced, and the surface reflects out of phase ($\Gamma\approx e^{j\pi}$). Second, when $X_{s}\gg\eta_{0}$ a ``high impedance surface" is produced. This is also called an ``artificial magnetic conductor'' since the surface acts similar to a perfect magnetic conductor (PMC), reflecting the wave back in phase ($\Gamma\approx e^{j2\pi}$).

In general, (\ref{eqn5}) can be written as
\begin{equation}\label{eqn6}
\vec{E}_r = E_{0}e^{-j\{(2\tan(\frac{X_{s}(x,y)}{\eta_{0}})+kd\}} \hat{y},
\end{equation}
and thus the phase of the reflected field is
\begin{equation}\label{eqn7}
\Phi(x,y) =-2\tan\left(\frac{X_{s}(x,y)}{\eta_{0}}\right)-kd.
\end{equation}
The first term in (\ref{eqn7}), the so-called ``reflection phase", is the phase difference between the incident and reflected fields at the surface of the array, and is solely due to the presence of the reflectarray. The second term is due to the propagation of the wave from the source to the reflecting surface. In the case of other kinds of incident waves, such as spherical waves, this term requires a different expression, which is of course a function of position on the reflectarray.

It is obvious from (\ref{eqn7}) that the properties of the surface, and hence $X_{s}$, can be changed in order to assign different phase shifts to different points on the array. As a result, one can form a desired radiation pattern by properly selecting the surface impedance at each point on the array. For instance, in order to obtain the maximum of the scattered field in the direction ($\theta$,$\phi$) shown in Fig. \ref{Sketch}, the reflected waves should constructively interfere at a point far away from the surface along those angles. This criterion leads to
\begin{equation}\label{eqn8}
\Phi(x,y)+k\sin\theta (x\sin\phi+y\cos\phi)-\Phi(0,0)=2n\pi,
\end{equation}
where $n$ is an integer number. Using (\ref{eqn7}), the phase criterion across the array can be written in terms of $X_{s}$ as
\begin{equation}\label{eqn9}
-2\left\{\tan\left(\frac{X_{s}(x,y)}{\eta_{0}}\right)-\tan\left(\frac{X_{s}(0,0)}{\eta_{0}}\right)\right\} +k\sin\theta=2n\pi.
\end{equation}

In microstrip reflectarrays, printed radiating elements are used to manipulate the surface impedance and hence generate the proper phase shift across the array needed to produce the desired beam shape. It is not possible to practically measure the reflection phase at each point on the surface of the reflectarray, so the average phase over a unit cell is used as a design parameter. The design of a single unit of the array is explained in the next section.

\section{Analysis of The Unit Cell}

Reflection phase, reflection efficiency, and the bandwidth of the element are the most important features of a unit cell used in a reflectarray antenna \cite{Huang2}. In this paper, the focus is mainly on the reflection phase properties of the unit cell. However, the reflection efficiency and the bandwidth will be briefly addressed.

\begin{figure}[!ht]
\begin{center}
\includegraphics[keepaspectratio = true, width = 1.5 in, clip = true]{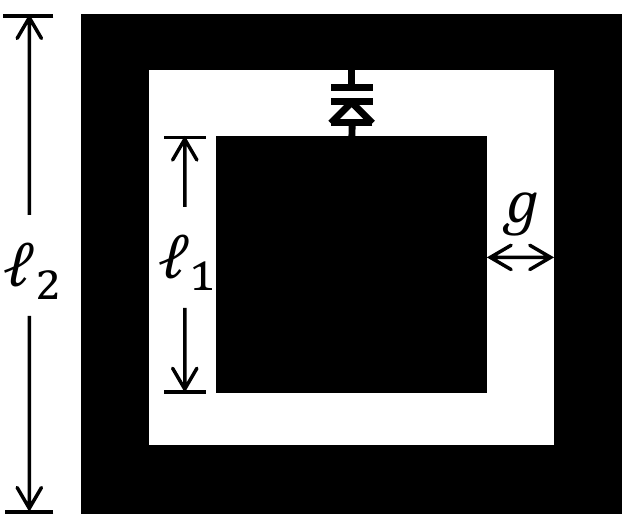}
\caption{Geometry of the unit cell.}  \label{unitcell}
\end{center}
\end{figure}

In order to achieve a dual-band, phase-adjustable element with a wide reflection phase range and a high reflection efficiency, an appropriate geometry must be considered. The proposed unit cell is shown in Fig. \ref{unitcell}. The geometry of the unit cell was inspired by a double square loop frequency selective surface (FSS), which perfectly blocks the incident electromagnetic wave at its two resonance frequencies \cite{Luo}. In this work, the inner loop is replaced  by a square patch since the square patch provides wider bandwidth. This improves the scattering properties of the unit cell. Hence, the proposed unit cell consists of a square patch centered in a square ring along with a varactor diode placed across the gap between the ring and the square patch to provide phase adjustability. A tunable dual-band unit cell based on square loops is studied in \cite{Bray} in the context of electromagnetic band-gap structures. The unit cell has the advantage of tuning each band independently. However, using six capacitors and inductors per unit cell increases the complexity of the structure.

The unit cell is analyzed in two ways -- with the aid of a commercial full wave solver (ANSYS HFSS), and by using an equivalent circuit model. In both cases it is assumed that the unit cell is printed on a square substrate of side $p$ (the periodicity of the unit cell) with a conducting ground on the reverse side. Although it is assumed here that the incident wave is a plane wave normal to the unit cell, so as to simplify the derivation of the equivalent circuit parameters, in principle the problem can be solved for TE or TM fields at arbitrary incidence.

\subsection{Full wave simulation}

The commercial full wave simulator ANSYS HFSS was used to analyze the scattering properties of the unit cell. Since waveguides are convenient for measuring the reflection and transmission coefficients, the unit cell was located inside a fictitious square waveguide and periodic boundary conditions used to model the array by infinitely replicating the unit cell. As a result, mutual interactions between the neighboring unit cells are taken into account.

The unit cell was designed to operate within S band (2.6-3.95~GHz) and F band (4.9-7.05~GHz). The geometry parameters were optimized in order to maximize the reflection coefficient and the phase change over the two bands. The periodicity of the unit cell, and thus the dimension of the square waveguide, was chosen to be 22~mm. The other parameters are $\ell_{2}=18$~mm, $\ell_{1}=12$~mm and $g=1$~mm. The substrate is 1.575~mm thick with a metal plane on the back, and the dielectric constant is set to unity. This last assumption is made in order to simplify the equations derived in the next section.

A varactor diode with a dynamic range of 0.1-0.6~pF is placed across the gap to achieve tunablity. Two key points have to be considered in order to properly incorporate the varactor diode into the unit cell. First, it is crucial that the diode offers an appropriate range of capacitances, otherwise it will dominate or be dominated by the intrinsic parameters of the unit cell. Second, the diode should be attached in a way that the polarity of the diode is aligned with the incident electric field. The presence of the diode breaks the symmetry of the structure. Therefore, unlike conventional double-square-loop elements, the unit cell no longer provides dual-polarized operations. However, the symmetrical structure of the unit cell does suppress cross polarization effects \cite{Hum2}. These two points will be visited in the investigation of the equivalent circuit model.

\begin{figure}[!ht]
\begin{center}
\includegraphics[keepaspectratio = true, width = 3.25 in, clip = true]{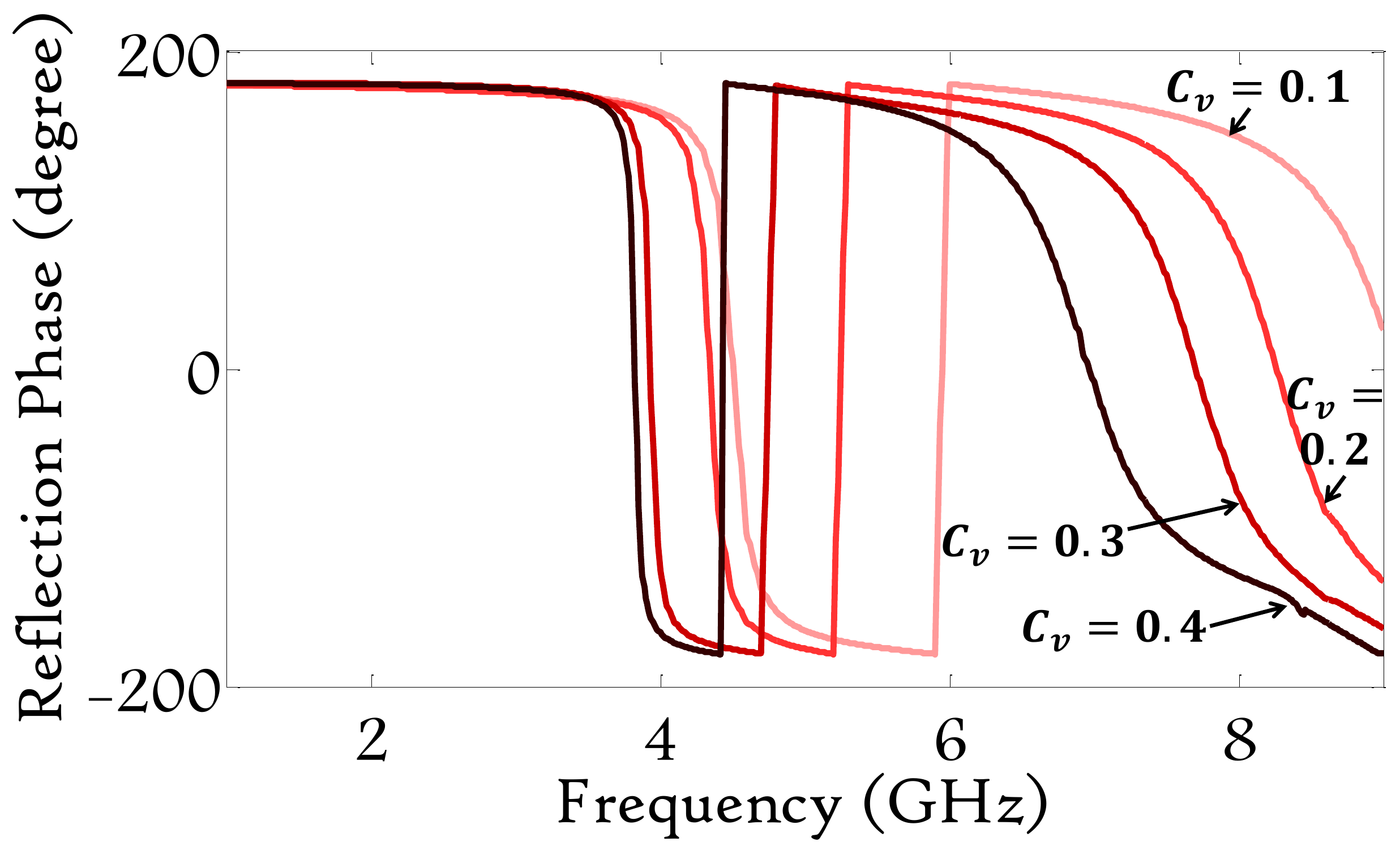}
\caption{Reflection phase for different varactor capacitance values, $C_{v}$(pF).}  \label{reflection phase HFSS ONLY}
\end{center}
\end{figure}

Fig. \ref{reflection phase HFSS ONLY} shows the reflection phase (\ref{eqn7}) as a function of frequency for different values of the varactor capacitance. The reflection phase has a range of almost 360$^{\circ}$, indicating high tuning ability. Moreover, two resonant frequencies can be observed. The higher resonant frequency is mainly due to the square patch, and the lower resonant frequency is largely due to the ring. This indicates that the array has the potential to steer the beam within two distinct frequency bands. Furthermore, the unit cell has the advantage of having stable performance over a wide range of incident angles 0-45$^{\circ}$ \cite{Lee}.

\begin{figure}[!ht]
\centering
\subfloat[]{\includegraphics[width=3.15in]{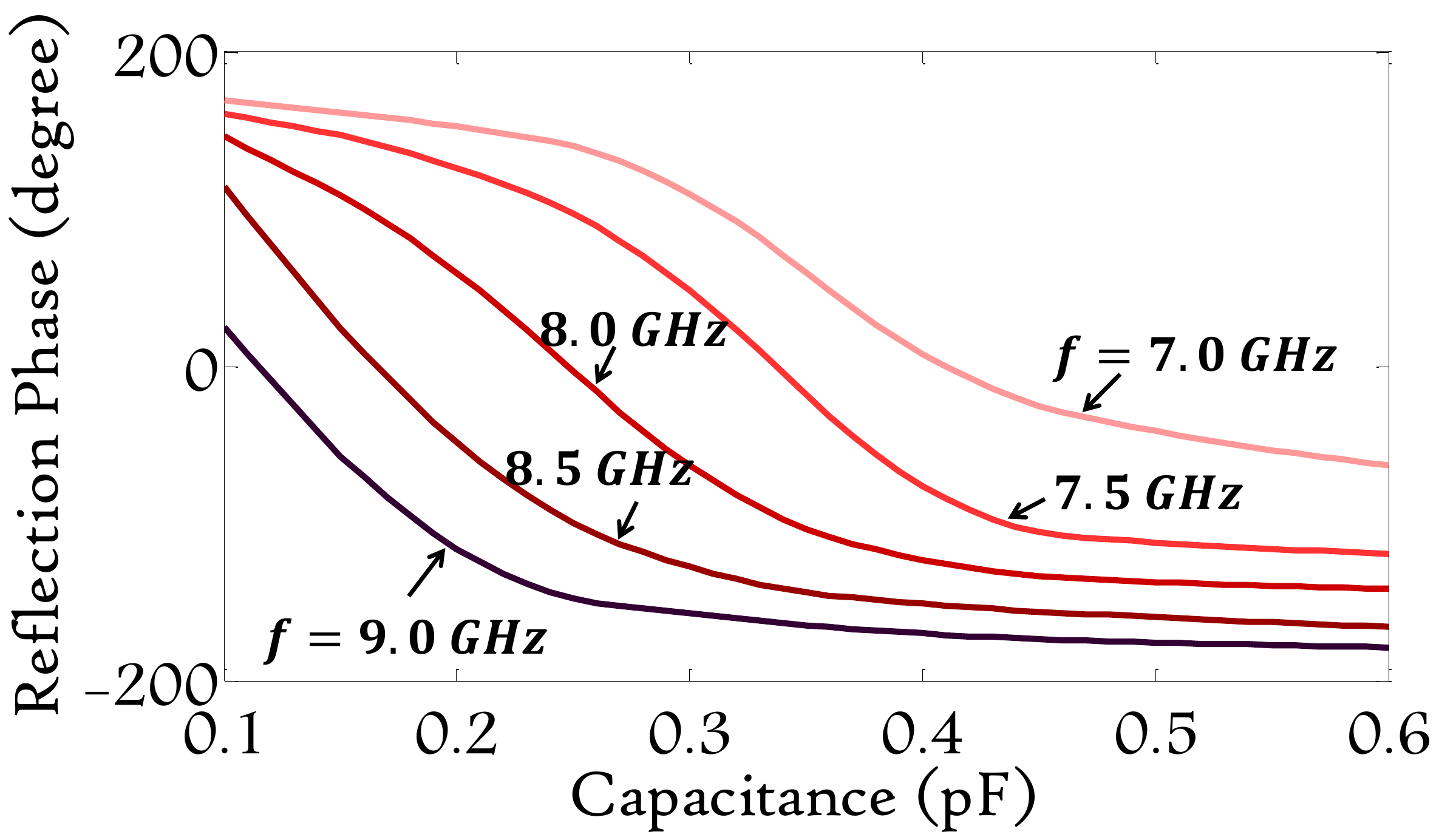}%
\label{refcurve}}
\hfil
\subfloat[]{\includegraphics[width=3.15in]{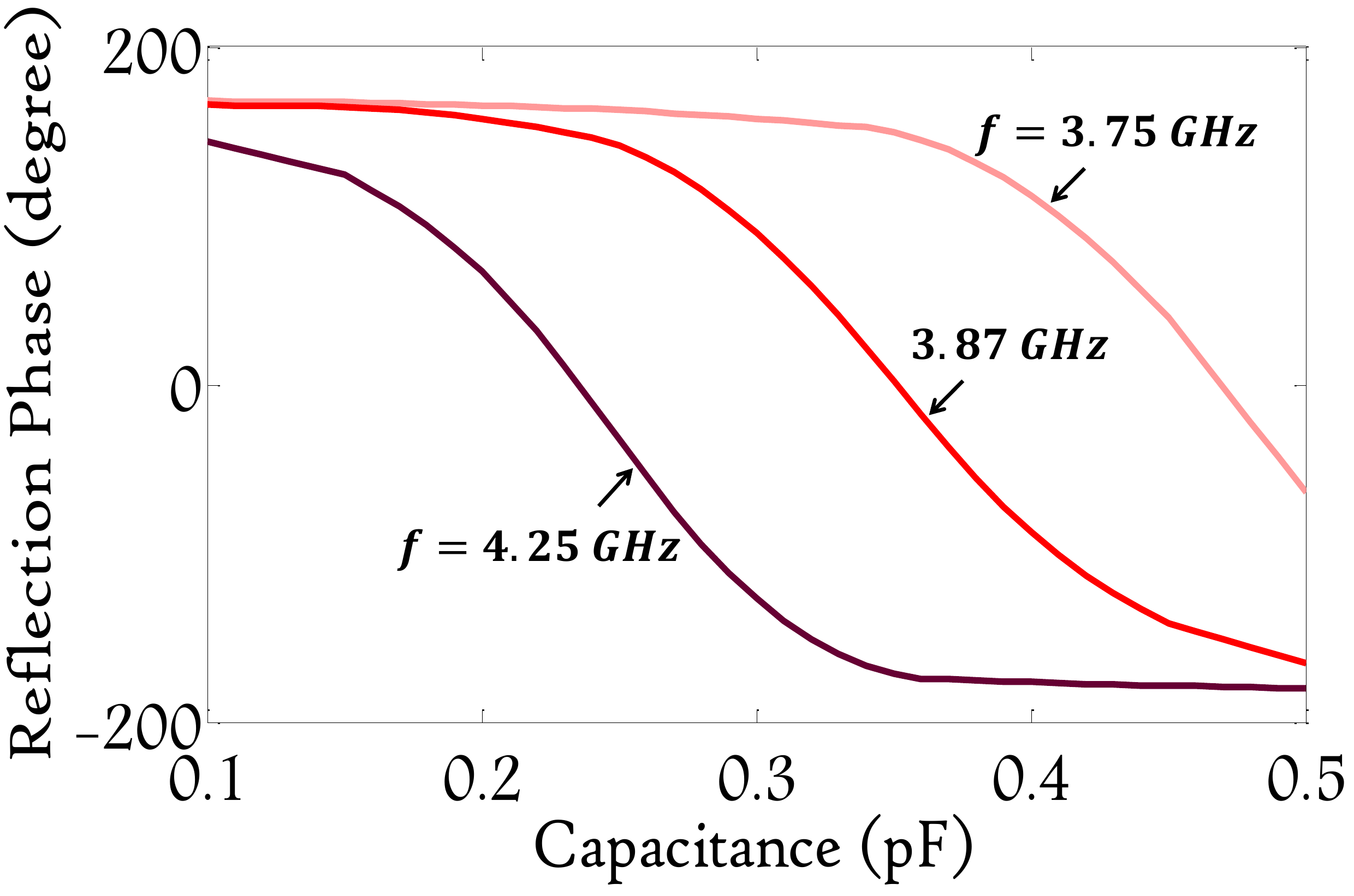}%
\label{figsecondcase2}}
\caption{Reflection phase at different frequencies in (a) the higher frequency band and (b)  the lower frequency band.}
\label{figsim2}
\end{figure}

Plots of the reflection phase as a function of the capacitance for both frequency bands are shown in Fig. \ref{figsim2}. It is clear that the phase reaches its maximum and minimum at the low and high values of the varactor capacitance ($C_{v}$) respectively. This suggests that the dynamic capacitance range of the diode has to be carefully chosen in order to maximize the phase shift produced by the unit cell. Notice that the parameters of the unit cell are designed in such a way that a single diode is capable of changing the reflection phase in the two bands simultaneously.

\begin{figure}[!ht]
\centering
\subfloat[]{\includegraphics[height=1.50in, width=3.10in]{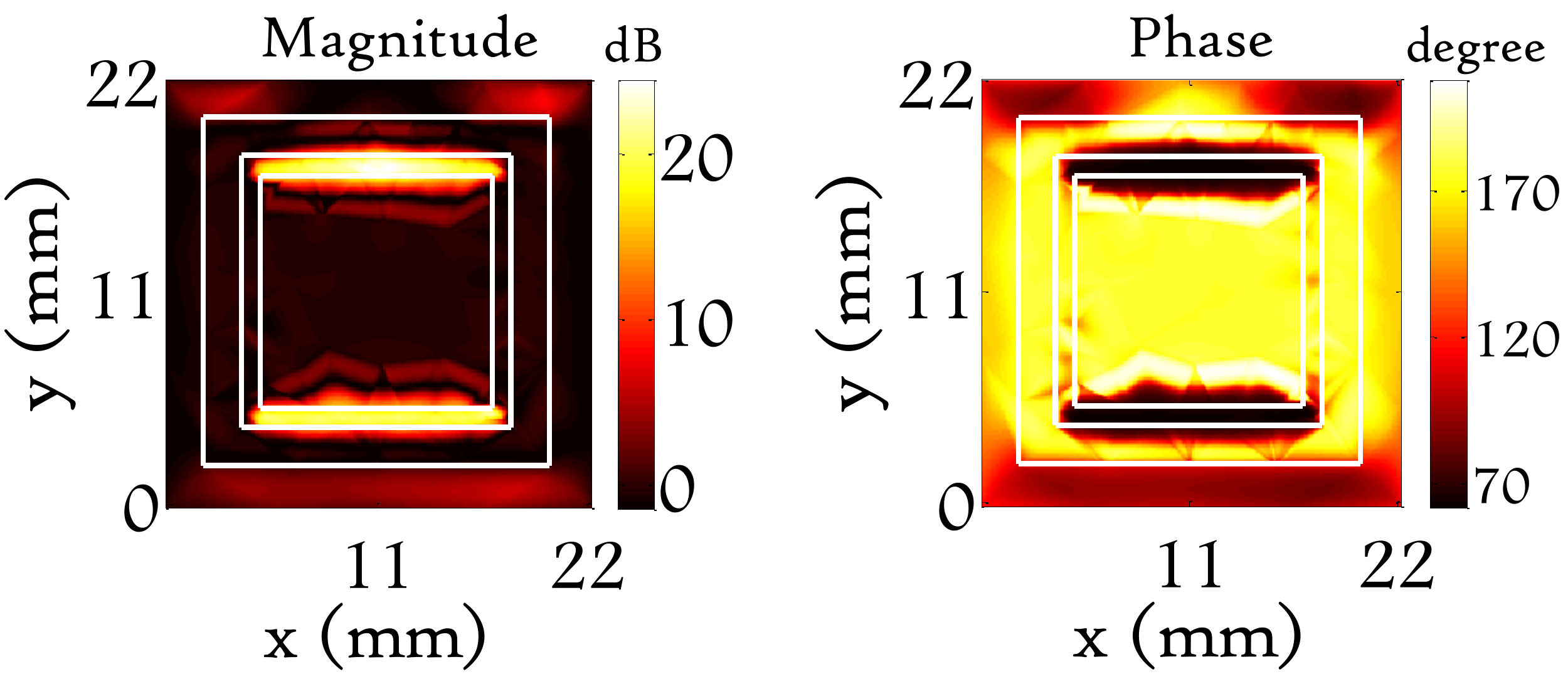}%
\label{phasegraphics1}}
\hfil
\subfloat[]{\includegraphics[height=1.50in, width=3.10in]{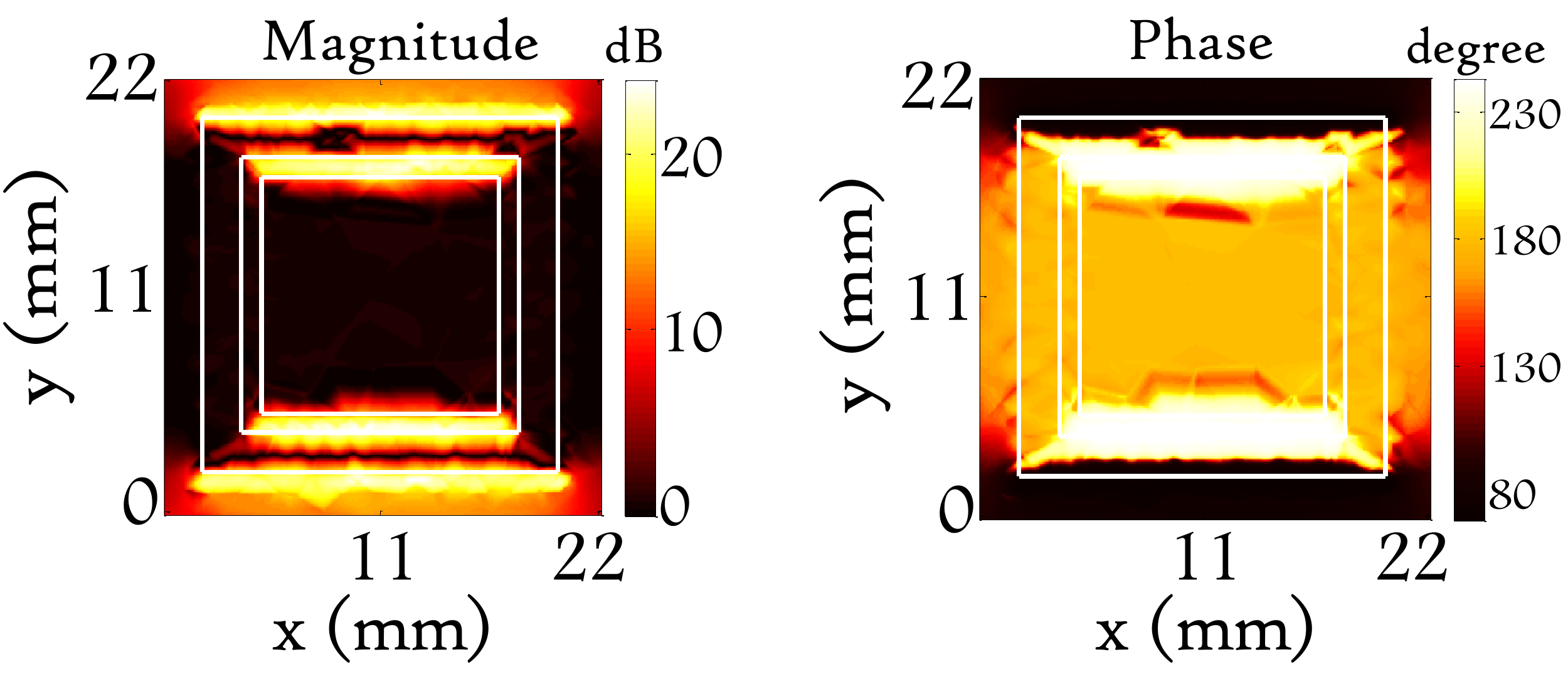}%
\label{phasegraphics2}}
\caption{Magnitude and phase of the reflection coefficient of the y component of the electric field on a plane 0.1 mm above the surface of the unit cell at (a) 8.0~GHz and $C_{v}$=0.2~pF and (b) 4.25~GHz and $C_{v}$=0.2~pF. The layout of the unit cell is shown with white lines.}
\label{phasegraphics_general}
\end{figure}

A good way to envision the role of the diode in changing the reflection phase is described in \cite{Rajagopalan} and adopted here. The magnitude and phase of the reflection coefficient of the electric field on a plane located 0.1~mm above the surface of the unit cell is shown in Fig. \ref{phasegraphics_general}. Note that the reflection phase given in Fig. \ref{reflection phase HFSS ONLY} and Fig. \ref{figsim2} can be interpreted as the phase of the average reflection coefficient at each point on this surface. To obtain $\Gamma$ on the surface, two simulations are required, one with the unit cell inside the waveguide in order to calculate the total electric field, and one without the unit cell in order to calculate the incident field. The scattered field is calculated in the post processing stage by subtracting the incident field from the total field, and finally $\Gamma$ is calculated with the aid of (\ref{eqn2}). The incoming wave is polarized along the $y$-axis (parallel to the diode polarity), and hence the magnitude and phase of the $x$ and $z$ components of the electric field are not considered. The magnitude and phase of the reflection coefficient of the $y$ component of the electric field at 8.0 GHz with $C_{v}=0.2$~pF are shown in Fig. \ref{phasegraphics_general}(a). As expected, the main contribution to the scattered field is from the two radiating edges of the square patch. Consequently, the reflection phase is greatly influenced by the phase at the radiating edges, and is approximately equal to it. In Fig. \ref{phasegraphics_general}(a) this phase is equal to $62^{\circ}$, which matches with the corresponding curve in Fig. \ref{figsim2}(a). Fig. \ref{phasegraphics_general}(b) shows the magnitude and phase of the reflection coefficient of the $y$ component electric field at 4.25~GHz with $C_{v}=0.2$~pF. It is obvious that the main contribution to the scattered field is from the two edges of the ring, so that again the phase of the two edges of the ring defines the final phase of the reflection coefficient. This phase is equal to $75^{\circ}$ and matches with Fig. \ref{figsim2}(b). Note that for both frequencies the phase of the reflection coefficient at the center of the unit cell, is $180^{\circ}$. This is because the edge effects are minimal at the center, thus the center behaves similar to a PEC.

\subsection{Equivalent Circuit Model}

One of the objectives of studying an FSS is the evaluation and optimization of the transmission and reflection coefficients of the incident wave with the help of an appropriate model \cite{Zhu}. Modeling an FSS using equivalent circuits is a well-known technique, which not only provides fast and accurate results but also reveals the physics of the problem. The first attempt in modeling an FSS with equivalent circuits can be found in \cite{Marcuvitz} where two simple cases of a field incident on an infinite array of metallic strips is considered. If the strips are parallel to the electric field, the surface can be modeled with a single inductor; if the strips are parallel to the magnetic field, the surface can be modeled with a single capacitor. Equivalent circuits of more complicated geometries such as square loops \cite{Langley}, gridded-square elements \cite{Lee}, double-square loops \cite{Lee}- \cite{Langley2} and Jerusalem crosses \cite{Campos} can be found by combining these two type of strips.

\begin{figure}[!ht]
\centering
\subfloat[]{\includegraphics[width=3.25in]{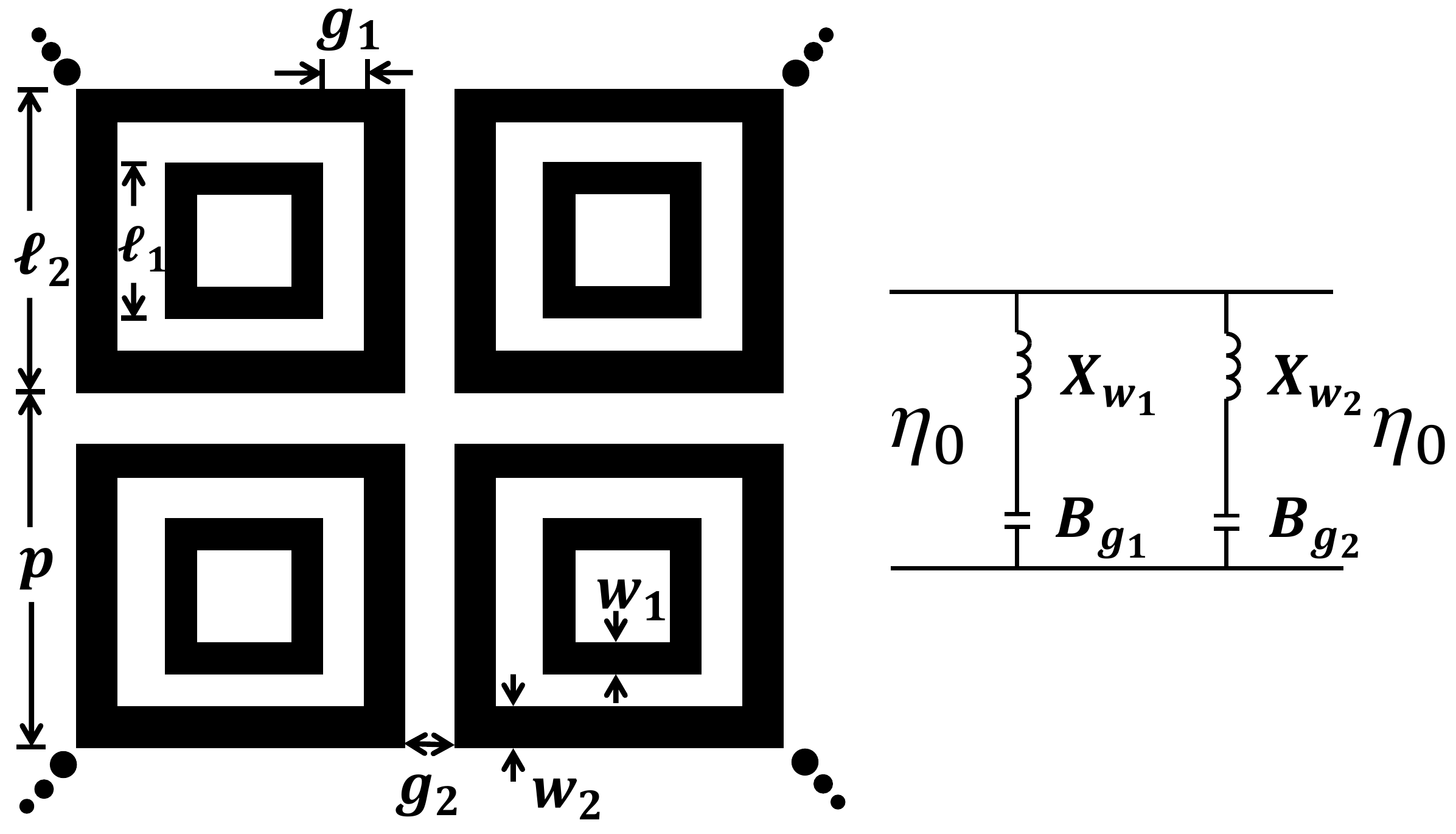}%
\label{equidoublesquareloop}}
\hfil
\subfloat[]{\includegraphics[width=3.25in]{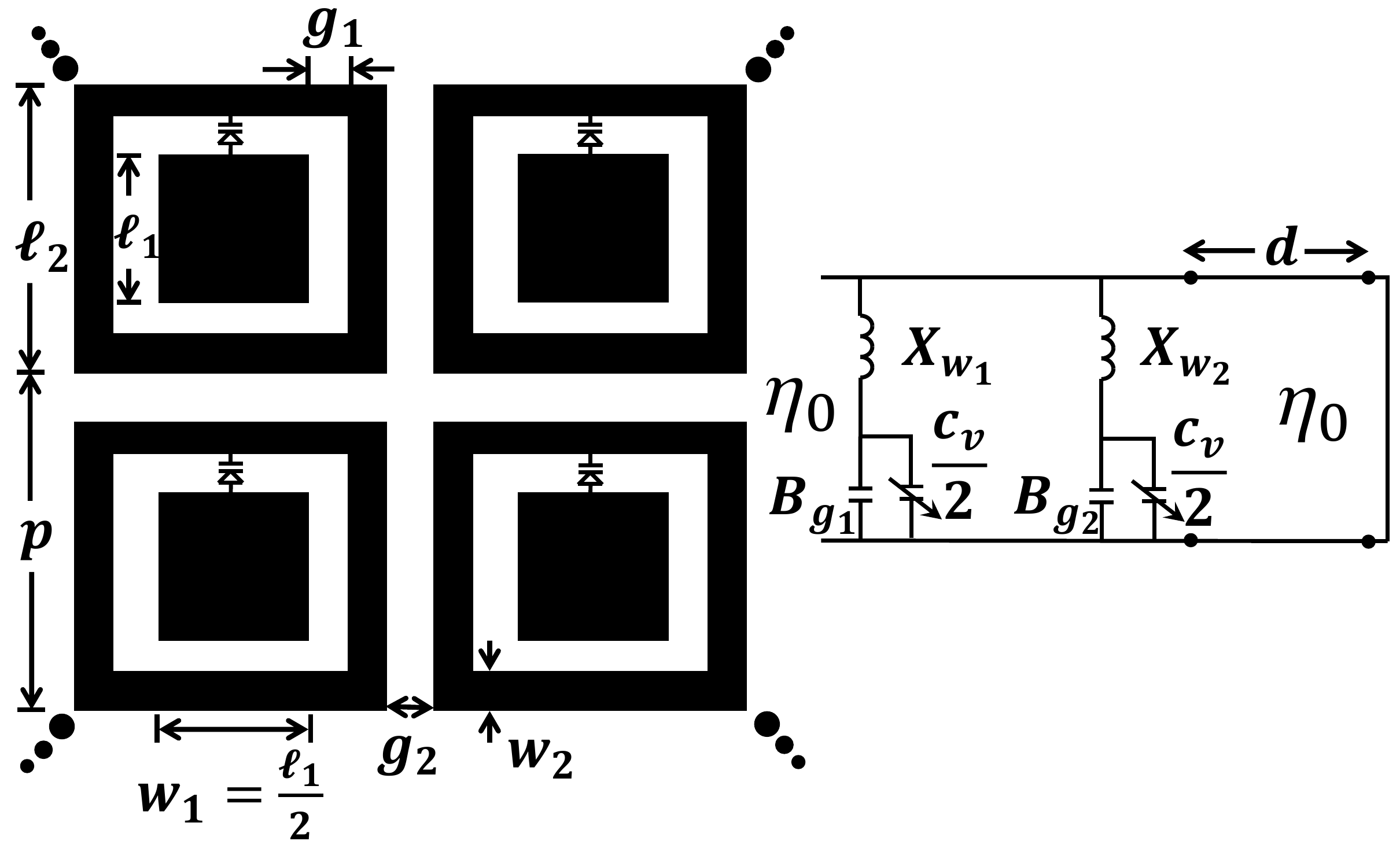}%
\label{equiunitcell}}
\caption{(a) Geometry of a double square loop and the equivalent circuit of an infite array of  double square loops. (b) Geometry of the proposed unit cell derived from a double square loop with $w_{1}=\ell_{1}/2$ and the corresponding equivalent circuit.}
\label{equiCirc}
\end{figure}

Fig. \ref{equiCirc}(a) shows the double loop geometry and its corresponding equivalent circuit \cite{Langley2}. The model consists of two shunt LC circuits representing the outer and inner loops.
It is easy to see that the limiting case of the double loop, described by $w_{1}=\ell_{1}/2$, lends itself well to the proposed unit cell, Fig. \ref{equiCirc}(b). One can modify the equivalent circuit of a double loop structure to achieve a model for the proposed unit cell as shown in the right panel of Fig. \ref{equiCirc}(b). For a TEM incident wave, the normalized values of the inductances and the capacitances of the equivalent circuit shown in Fig. \ref{equiCirc}(a) are given by \cite{Langley2}
\begin{equation}\label{eqn10}
\begin{aligned}
\frac{X_{w_{1}}}{\eta_{0}}&= \frac {\omega L_{w_{1}}}{\eta_{0}}=X_{1}\frac{\ell_{1}}{p},\\
\frac{X_{w_{2}}}{\eta_{0}}&=\frac {\omega L_{w_{2}}}{\eta_{0}}=2\frac{X_{2}X_{3}}{X_{2}+X_{3}}\frac{\ell_{2}}{p},\\
B_{g_{1}}\eta_{0}&=\omega C_{g_{1}}\eta_{0}=\frac{B_{1}B_{2}}{B_{1}+B_{2}}\frac{\ell_{1}}{p},\\
B_{g_{2}}\eta_{0}&=\omega C_{g_{2}}\eta_{0}=\frac{3}{4}B_{2}\frac{\ell_{2}}{p}.
\end{aligned}
\end{equation}
Here $X_{1}$, $X_{2}$, $X_{3}$, $B_{1}$ and $B_{2}$ in (\ref{eqn10}) are given by
\begin{equation}\label{eqn11}
\begin{aligned}
X_{1}&= F(p,2w_{1},\lambda),\\
X_{2}&= F(p,w_{2},\lambda),\\
X_{3}&= F(p,w_{1},\lambda),\\
B_{1}&= 4F(p,g_{1},\lambda),\\
B_{2}&= 4F(p,g_{2},\lambda),
\end{aligned}
\end{equation}
where $F(p,x,\lambda)$ has a general form of
\begin{equation}\label{eqn12}
F(p,x,\lambda)=\frac{p}{\lambda} \left[\ln \csc \left(\frac{\pi x}{2p}\right)+G(p,x,\lambda)\right].
\end{equation}
Here $G(p,x,\lambda)$ is a correction factor given by
\begin{equation}\label{eqn13}
\begin{split}
G(p,x,\lambda)=\\
\frac{1}{2}\frac{(1-\beta^2)^2[2C(1-\beta^2/4)+4C^2\beta^2]}{(1-\beta^2/4)+2C\beta^2(1+\beta^2/2-\beta^4/8)+2C^2\beta^6}
\end{split}
\end{equation}
where
\begin{equation}\label{eqn14}
\indent \beta=\sin\left(\frac{\pi x}{2p}\right)
\end{equation}
and
\begin{equation}\label{eqn15}
\indent C=\frac{1}{\sqrt{1-(p/\lambda)^2}}-1
\end{equation}
The model accurately predicts the behavior of a double loop as long as $w_{1}$, $w_{2}$, $g_{1}$, and $g_{2}$ are all much less than $p$, and $p<\lambda$.

The values of the inductances and capacitances of the unit cell equivalent circuit can be found using (\ref{eqn10})-(\ref{eqn15}). In addition, the effect of the substrate, the metal plate on the back, and the varactor diode should be taken into account. The substrate is treated as a transmission line with a length equal to the thickness of the substrate, while the ground plate provides a short circuit at the end of the line. Hence the impedance seen at the surface due to the substrate is
\begin{equation}\label{eqn16}
\indent Z_{1}=j\eta_{0}\tan(\beta d),
\end{equation}
where $\eta_{0}$ is the free space impedance, $\beta=2\pi f \sqrt{\mu_{0}\epsilon_{0}}$ is the phase constant at frequency $f$, $\epsilon_{0}$ and $\mu_{0}$ are the free space permittivity and permeability respectively, and $d$ is the thickness of the substrate.
The effect of the varactor diode is taken into account by assuming that the diode almost equally shifts the two resonant frequencies, and thus it equally affects the outer loop and the patch. It is clear that the diode polarity should be along the incident electric field. This configuration minimally perturb the intrinsic inductances of the unit cell. The surface impedance (see equations 3 and 4) of the array can be calculated as
\begin{equation}\label{eqn17}
\indent Z_{s}=\eta_{0}/Y_{s}=\eta_{0}/(Y_{1}+Y_{2}+Y_{3})
\end{equation}
The quantity $Y_{s}$ is the normalized surface admittance. $Y_{1}$, $Y_{2}$ and $Y_{3}$ are the normalized admittances corresponding to the substrate (equation 16), the ring and the patch respectively, and are given by
\begin{equation}\label{eqn18}
\begin{aligned}
\indent Y_{1}&=1/Z_{1},\\
\indent Y_{2}&=\frac{j(B_{g_{2}}+B_{C_{v}}/2)}{1-X_{w_{2}}(B_{g_{2}}+B_{C_{v}}/2)},\\
\indent Y_{3}&=\frac{j(B_{g_{1}}+B_{C_{v}}/2)}{1-X_{w_{1}}(B_{g_{1}}+B_{C_{v}}/2)},
\end{aligned}
\end{equation}
where $B_{C_{v}}=\eta_{0}wC_{v}$ is the normalized reactance of the capacitor.

\begin{figure}[!ht]
\begin{center}
\includegraphics[keepaspectratio = true, width = 3.25 in, clip = true]{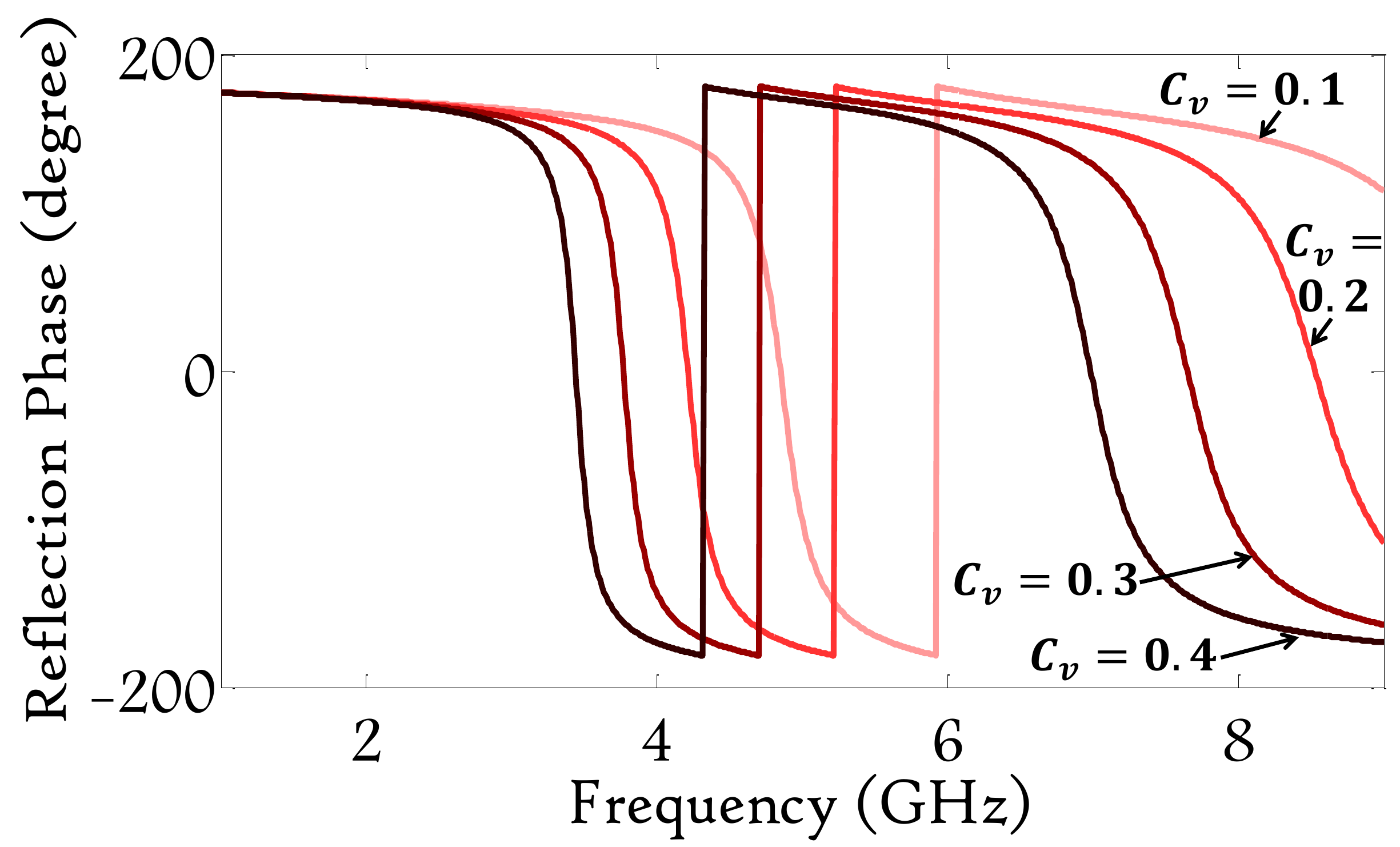}
\caption{Reflection phase curves for different capacitance values, $C_{v}$(pF), predicted by the equivalent circuit model.}  \label{reflection phase Eq Vs HFSS}
\end{center}
\end{figure}

Notice that since there is no lossy element in the circuit, $Z_{s}$ is purely reactive. Once the surface impedance is known, the phase of the reflection coefficient can be calculated as shown in (\ref{eqn3}),
\begin{equation}\label{eqn19}
\indent \phi=\Im\left\{\ln\frac{Z_{s}-\eta_{0}}{Z_{s}+\eta_{0}}\right\}
\end{equation}
The reflection phase predicted by the circuit is shown in Fig. \ref{reflection phase Eq Vs HFSS}. The parameters are the same as in the previous section. A good agreement can be seen between the phase predicted by the circuit (Fig. \ref{reflection phase Eq Vs HFSS}) and predicted by HFSS (Fig. \ref{reflection phase HFSS ONLY}). Note that with the dimensions mentioned, the condition $w_{1}\ll p$ still holds. It is seen that the addition of $C_{v}$ shifts the resonance frequency of the structure.

It is worth mentioning that, in principle, the effect of dielectric substrates can be taken into account in various ways. In \cite{Munk}-\cite{Costa} the frequency axis is scaled by a factor of $1/\sqrt{\epsilon_{r}}$. In \cite{Luo} the effect of the dielectric properties is incorporated in the equivalent susceptances given in (\ref{eqn11}). Although the dielectric properties have not been considered so far, in the next section where simulation and experimental results are compared the substrate properties are taken into account.

\subsection{Measurements}

\begin{figure}[!ht]
\centering
\subfloat[]{\includegraphics[width=3.45in, height=1.25in]{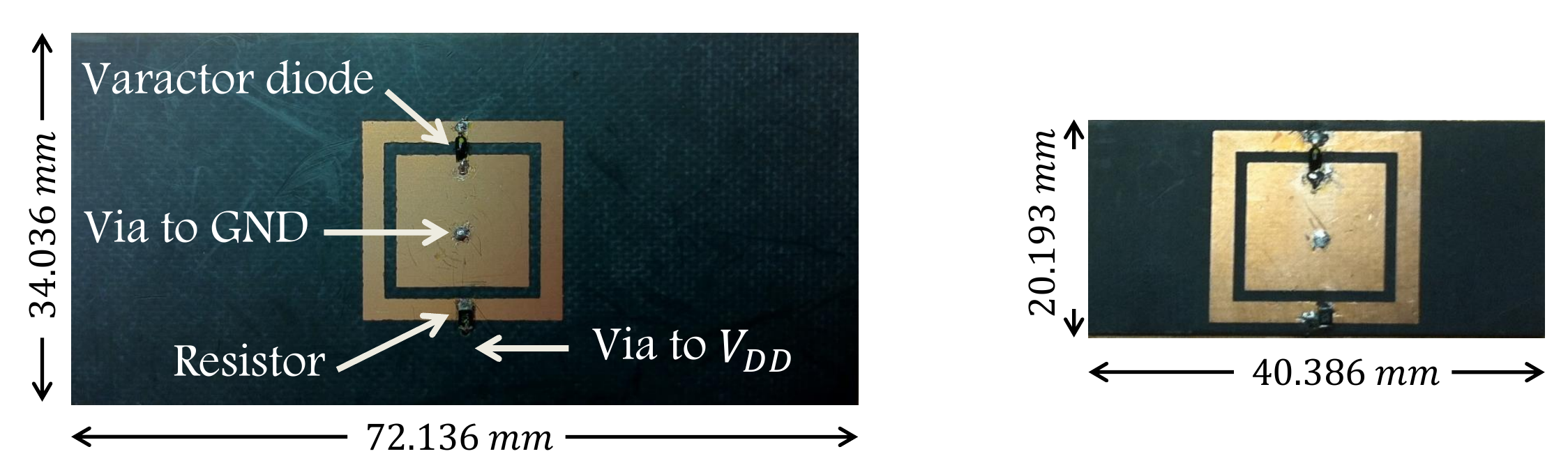}%
\label{expunitcells}}
\hfil
\subfloat[]{\includegraphics[width=3.20in, height=1.65in]{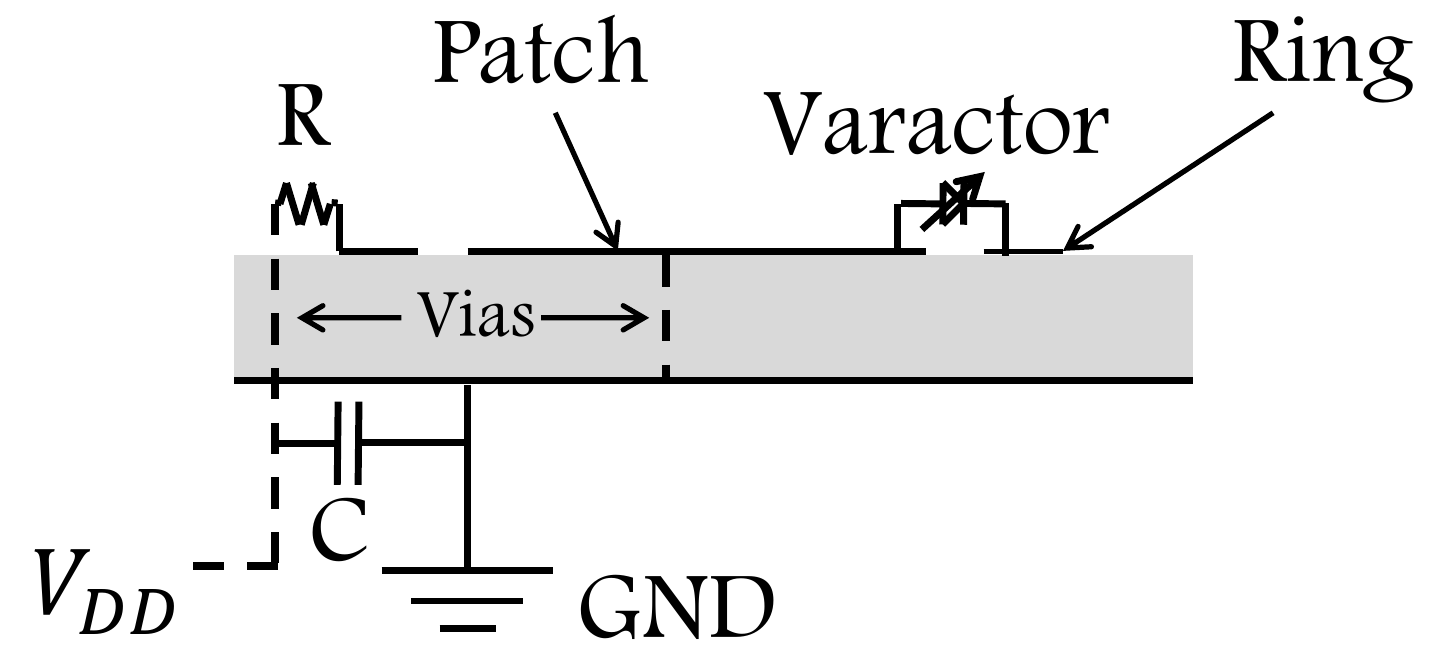}%
\label{exuiunitcell}}
\caption{(a) Unit cells. Left: S-band. Right: F-band. (b) Side-view schematic of the biasing circuit.}
\label{unitcellexperimental}
\end{figure}

Since the higher and lower resonance frequencies of the structure lie in the F (4.9-7.05~GHz) and S (2.6-3.95~GHz) frequency bands respectively, two waveguides were used to experimentally validate the simulated results. Two prototypes based on the aperture sizes of the waveguides were fabricated to completely fill the waveguide cross sections; see Fig. \ref{unitcellexperimental}(a). The unit cells with the dimensions provided in section~\MakeUppercase{\romannumeral 3}~A, were fabricated on a 1.575~mm thick RT/Duroid 5880 substrate with $\epsilon_r$ of 2.2 and $\tan\delta=0.0009$. The biasing circuit for the varactor diode, shown in Fig. \ref{unitcellexperimental}(b), consists of a resistor R=10~M$\Omega$ and a capacitor C=1000~pF, and prevents the structure from coupling to the bias line, bypassing the RF noise from the power source. Vias are used in order to preserve a high reflection efficiency. This eliminates the need to use wires and circuitry on the front side of the unit cell, which would perturb the scattered field. The appropriate placement of the vias and the biasing components was determined by investigating the surface current distribution on the structure using HFSS. The regions with relatively low surface currents were considered and different combinations of the vias located in these areas were implemented and compared. The best configuration was chosen to produce minimum impact on the performance of the resonant structure. One via close to the center of the square patch is connected to the ground plane, while another via is placed slightly below the bottom edge of the square ring, thus creating enough space for mounting the resistor on the ring.

A GaAs hyperabrupt varactor diode (Aeroflex Metelics MGV100-20) with a practical dynamic tuning range of $C_{v}=0.18-2.0$~pF (corresponding to the voltage range $28-0.5$~V) is used, since the diodes have both low parasitic parameters and low power dissipation. The series resistance of the diode is 3~$\Omega$ and the parasitic inductance is 0.4~nH.

In order to properly analyze the behavior of the unit cell inside the two waveguides, one has to consider two key differences from the model described in the previous sections. First, since the S-band and F-band unit cells have different periodicities, the equivalent infinite arrays generated by imaging into the waveguide walls have different unit cell spacings. This spacing is also different from the initial design ($p$ in section \MakeUppercase{\romannumeral 3} A and B). This alters the scattering properties, shifts the resonance frequencies, and as a result the reflection phase curves are shifted. Second, real waveguides do not support TEM modes. The dominant propagating mode is TE$_{10}$, which is a superposition of two plane waves bouncing between the waveguide walls, at angle $\theta$ with respect to the main symmetry axis of the waveguide. The angle $\theta$ for the TE$_{10}$ mode is given as \cite{Hannan}
\begin{equation}\label{eqn20}
\indent \theta=\sin^{-1}\left(\frac{\lambda_0}{2a}\right)
\end{equation}
where $\lambda_0$ is the free space wavelength and $2a$ is the cutoff wavelength of the TE$_{10}$ mode, where $a$ is the dimension of the longer side of the rectangular waveguide cross section. For the S-band and F-band waveguides, at their central frequencies this angle is $\theta=39.38^{\circ}$ and $\theta=38.40^{\circ}$, respectively. These deviations from the original model considered in the previous section have to be incorporated and the simulation model has to be altered correspondingly before comparing the simulation results with the measurements.

\begin{figure}[!ht]
\centering
\subfloat[]{\includegraphics[width=3.25in]{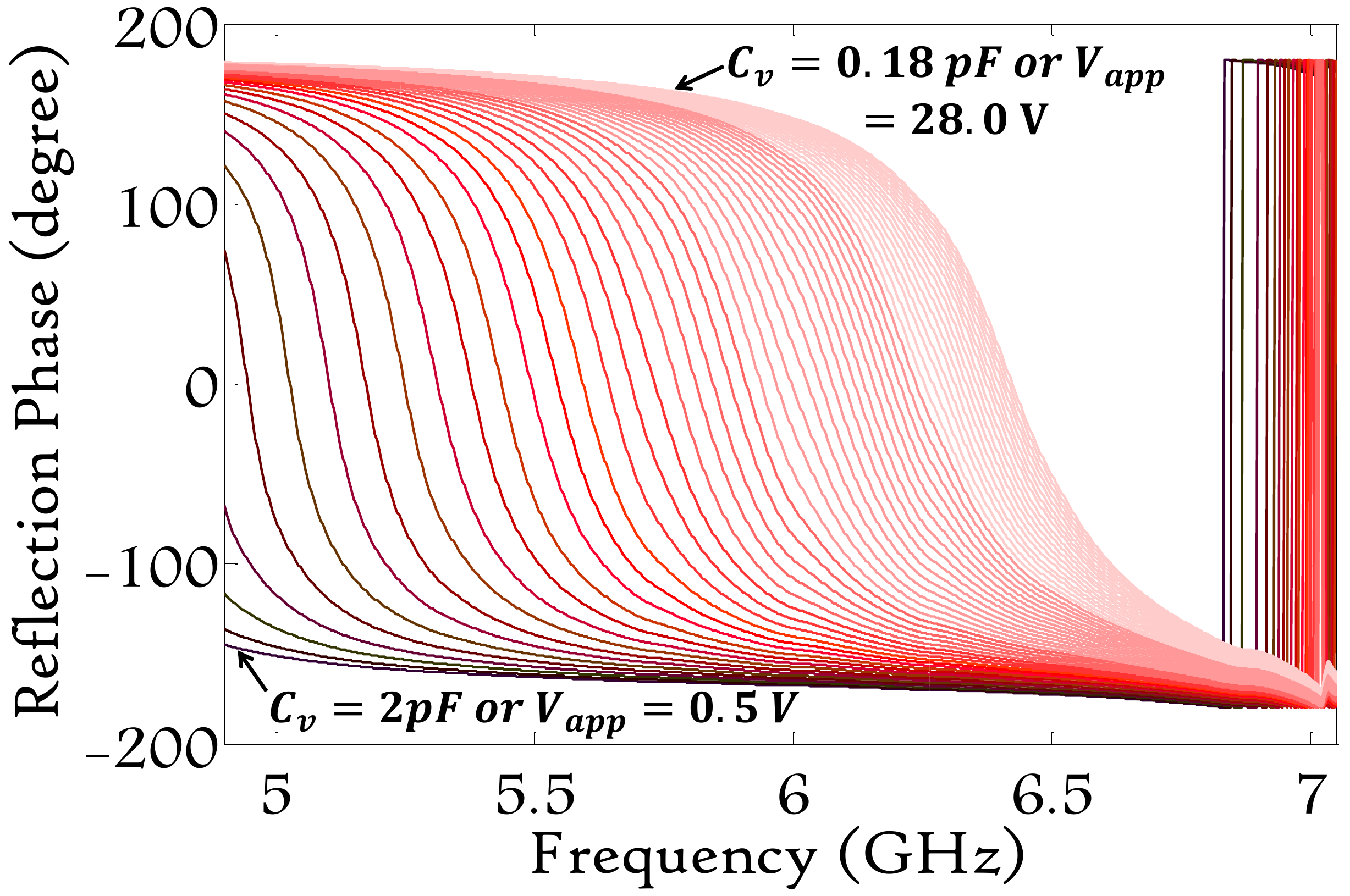}%
\label{Fbandunitcell}}
\hfil
\subfloat[]{\includegraphics[width=3.25in]{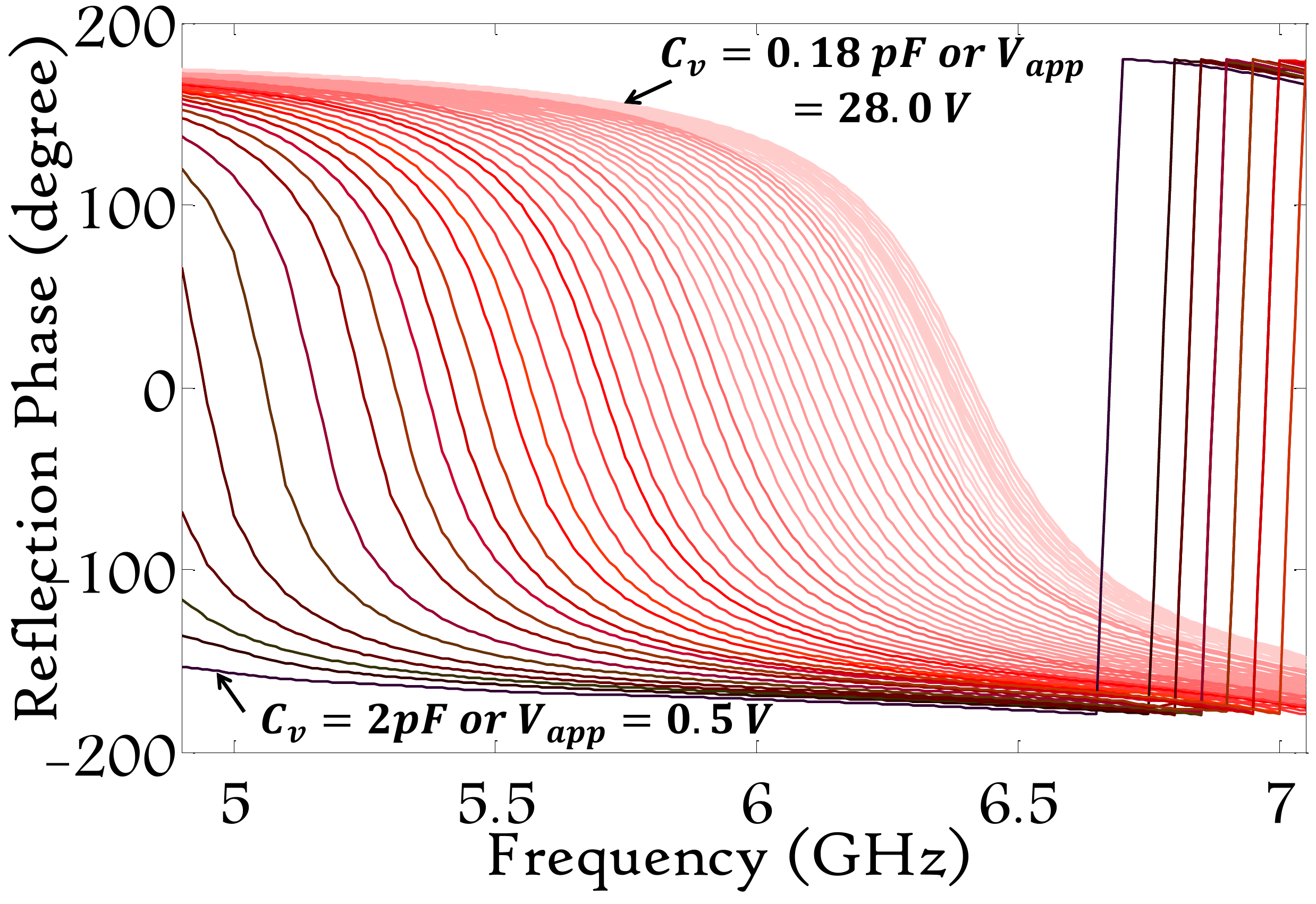}%
\label{FbandHFSSVSModeling}}
\caption{Reflection phase of the F-band unit cell as a function of frequency for different voltages (or equivalently $C_{v}$) (a) experiment (b) simulation.}
\label{Fband}
\end{figure}

Fig. \ref{Fband}(a) shows the measured reflection phase of the F-band unit cell with a varactor diode biased under various voltages. The applied voltage varies from 0.5~V to 28~V (with 0.5~V increments) corresponding to a capacitance range of 2.0~pF to 0.18~pF.  Simulations of the unit cell under the same capacitance conditions as the experiment were performed, and the phase responses are shown in Fig. \ref{Fband}(b). It is seen that good agreement of the phase response between measurement and simulation is achieved, and the unit cell is able to provide excellent phase shift capability over a wide frequency range. A maximum phase range of 335$^{\circ}$ is achieved experimentally at 5.30~GHz, and acceptable phase tuning ability over a wide frequency band of 5.0~GHz to 6.25~GHz can still be obtained. Note that the reflection phases are different from the unit cell with a TEM incident wave, which are shown in Fig. \ref{reflection phase HFSS ONLY} and Fig. \ref{reflection phase Eq Vs HFSS}. This suggests that the oblique incidence together with larger periodicity has shifted the resonance frequency.

\begin{figure}[!ht]
\centering
\subfloat[]{\includegraphics[width=3.25in]{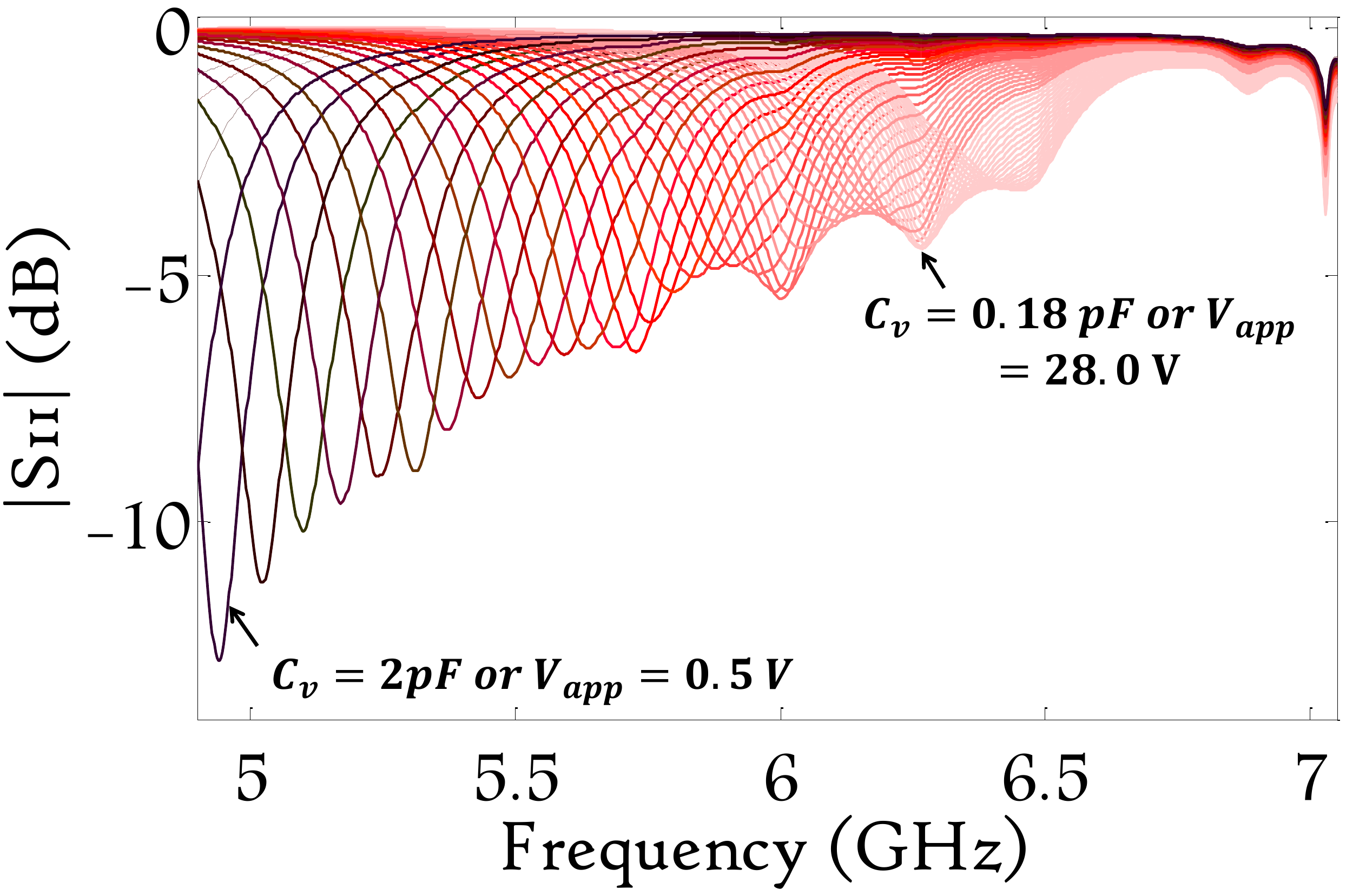}%
\label{FbandunitcellS11}}
\hfil
\subfloat[]{\includegraphics[width=3.20in]{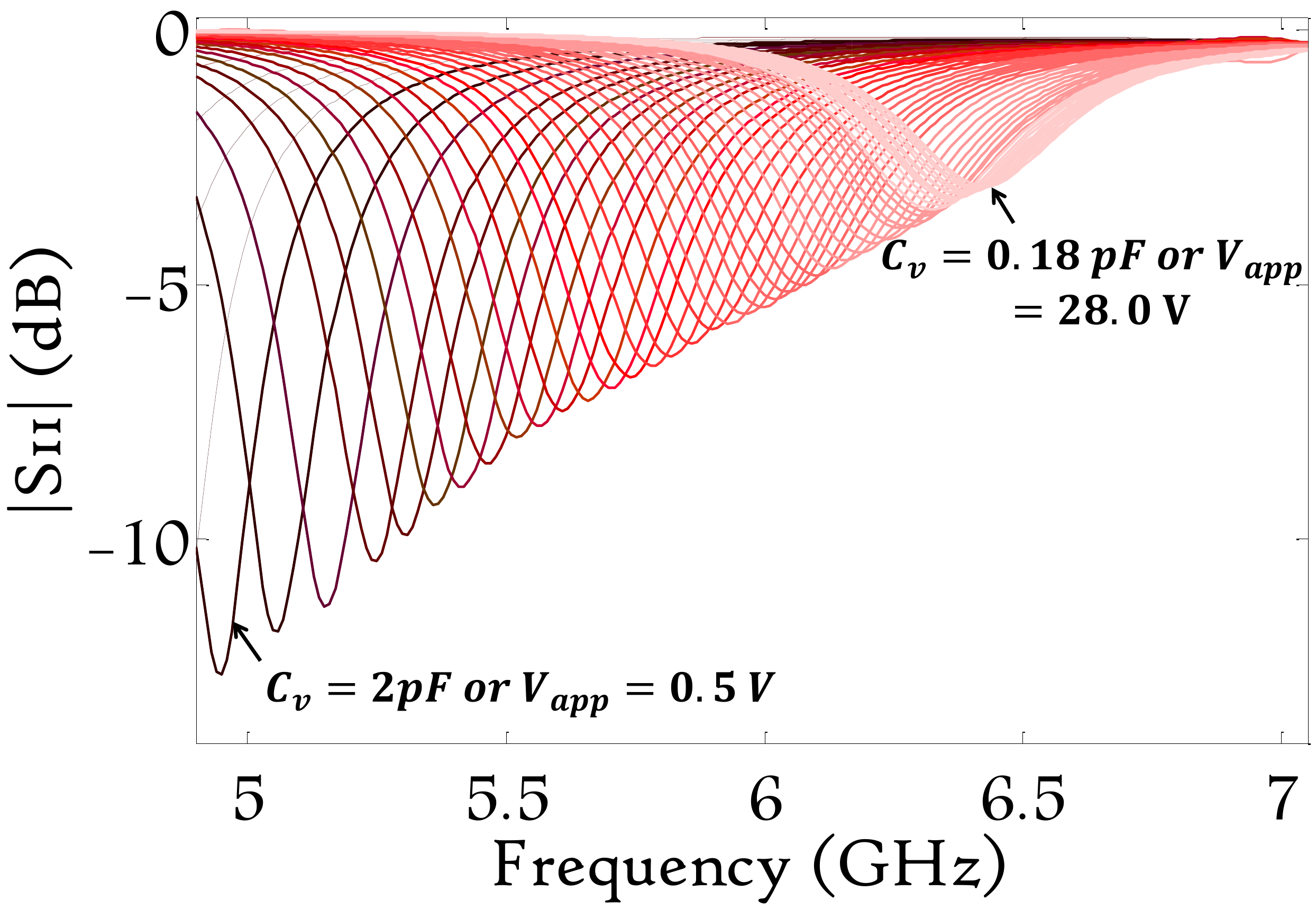}%
\label{FbandHFSSVSModelingS11}}
\caption{Magnitude of the reflection coefficient of the F-band unit cell as a function of freqeuncy for different voltages (a) experiment (b) simulation.}
\label{FbandS11}
\end{figure}

Fig. \ref{FbandS11}(a) and Fig. \ref{FbandS11}(b) show the measured and simulated magnitude of the reflection coefficient respectively, with the same biasing configuration as in Fig. \ref{Fband}. As the simulation predicts, the measured return loss of the unit cell decreases as the voltage increases.  For an applied voltage larger than 8~V, a return loss that is less than 5~dB can be achieved, which implies a potential high reflection from the unit cell.

\begin{figure}[!ht]
\centering
\subfloat[]{\includegraphics[width=3.25in]{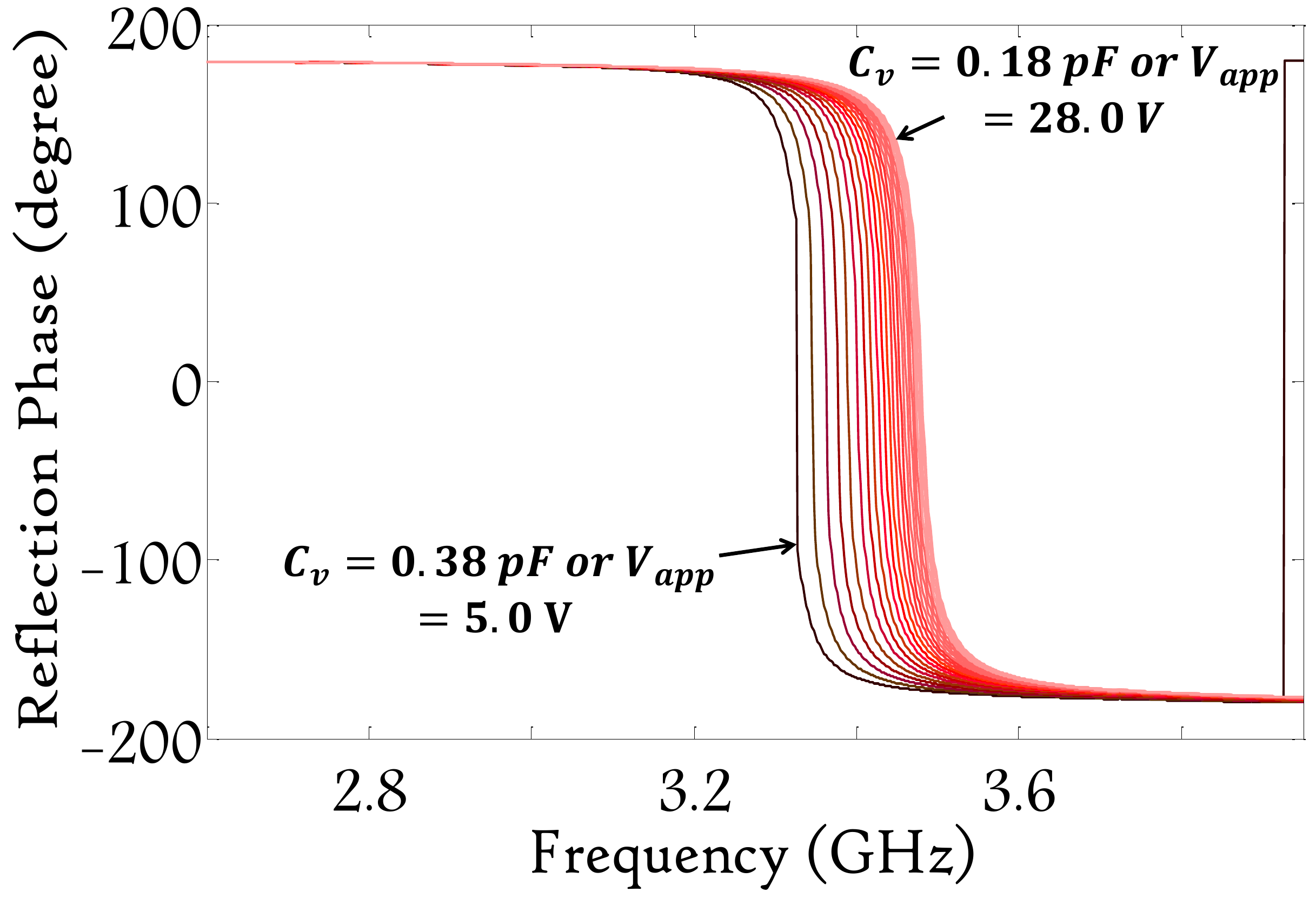}%
\label{Sbandunitcell}}
\hfil
\subfloat[]{\includegraphics[width=3.25in]{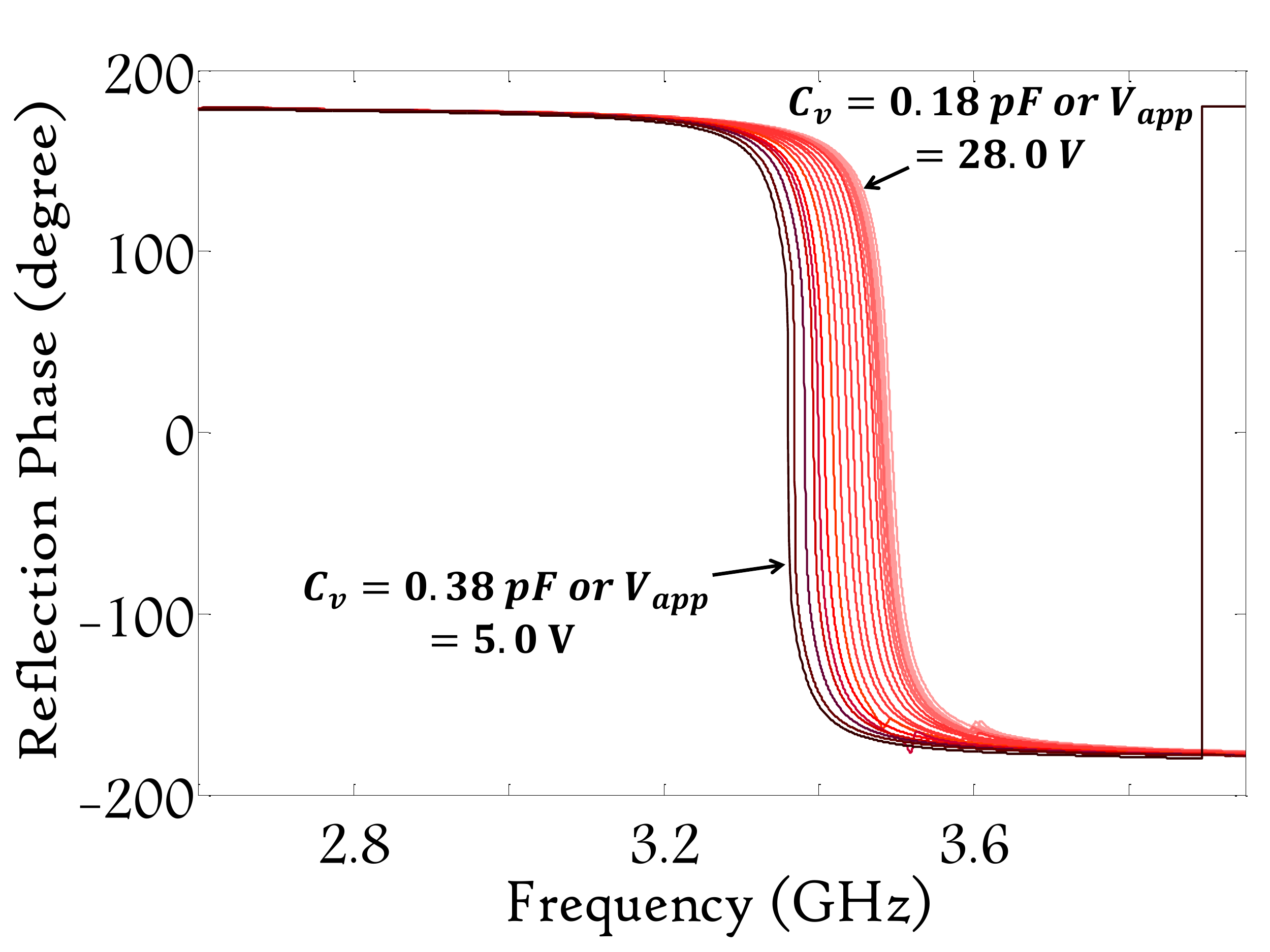}%
\label{SbandHFSSVSModeling}}
\caption{Reflection phase of the S-band unit cell as a function of frequency for different voltages (or equivalently $C_{v}$) (a) experiment (b) simulation.}
\label{Sbandph}
\end{figure}

\begin{figure}[!ht]
\centering
\subfloat[]{\includegraphics[width=3.25in]{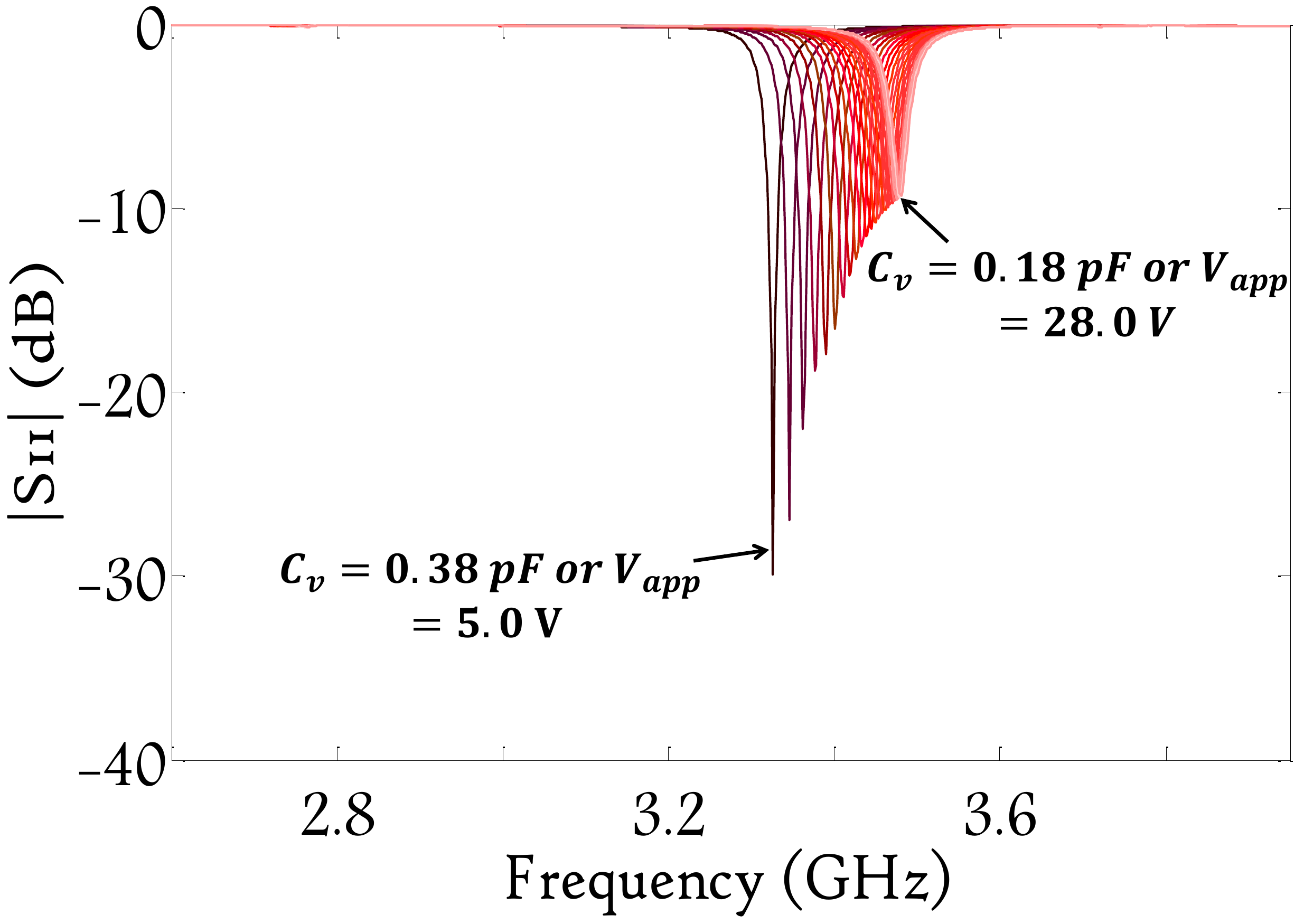}%
\label{Sbandunitcells11}}
\hfil
\subfloat[]{\includegraphics[width=3.3in]{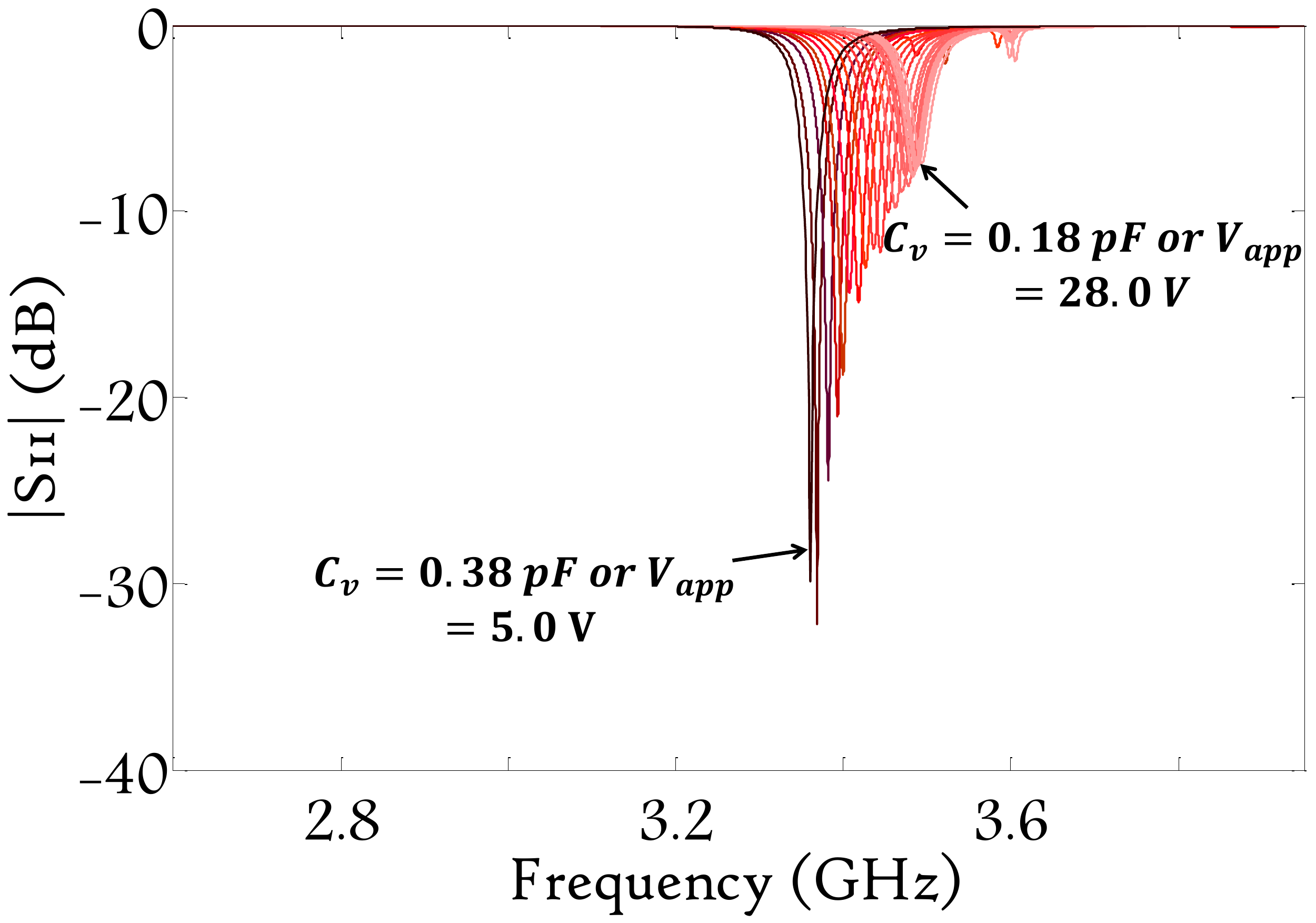}%
\label{SbandHFSSVSModelings11}}
\caption{Magnitude of the reflection coefficient of the S-band unit cell as a function of freqeuncy for different voltages (a) experiment (b) simulation.}
\label{Sbandmag}
\end{figure}

Both experimental and simulated results of phase and return loss of the S-band unit cell are provided in Fig. \ref{Sbandph} and Fig. \ref{Sbandmag}. It can be seen from Fig. \ref{Sbandph} that parallel results between simulation and experiment are achieved and a maximum phase range of 340$^{\circ}$ is reached at 3.49~GHz . Note that the phase responses are for a voltage range of 5~V to 28~V (with 1~V increments) or an equivalent capacitance range of 0.38~pF to 0.18~pF.  Reflection phase for a lower biasing voltage or higher capacitance can still be obtained. However, discontinuous variations of the phase responses are observed, which make the unit cell under these configurations not suitable for beam steering purposes. This issue can be explained as follows. From (\ref{eqn12}), the value of different elements in the equivalent circuit of Fig. \ref{equiCirc}(b) are proportional to $1/\lambda$, and thus the intrinsic capacitance of the unit cell is lower in the S-band. As a result, for voltages lower than 5~V, the capacitance of the varactor diode dominates the intrinsic capacitance of the unit cell, and therefore interrupts the continuous phase variations of the unit cell. In addition, in contrast with the wide frequency range at F-band, the operating bandwidth for this unit cell is relatively narrow. This is due to the narrow bandwidth nature of the square ring structure.

Fig. \ref{Sbandmag} shows the magnitude of the reflection coefficient. Reasonable agreement is seen between simulation and measurement. Note that the S-band unit cell exhibits a lower reflection than the F-band unit cell. This is because of the lossier nature of the ring compared to the square patch. In addition, the larger periodicity of the S-band unit cell affects the efficiency negatively.

\section{Fabricated Reflectarray}

\begin{figure}[!ht]
\centering
\subfloat[]{\includegraphics[width=3.55in, height=1.7in]{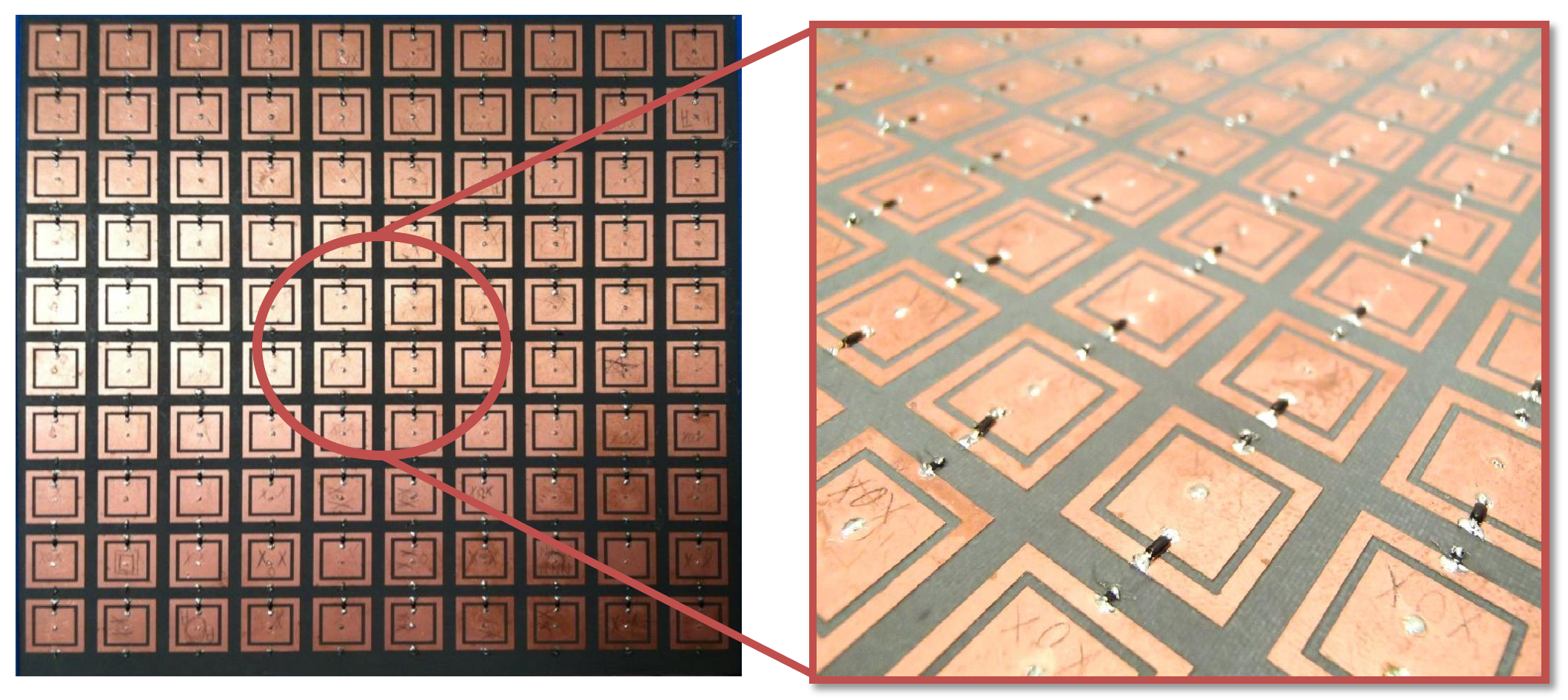}%
\label{reftpic}}
\hfil
\subfloat[]{\includegraphics[width=3.7in, height=2.5in]{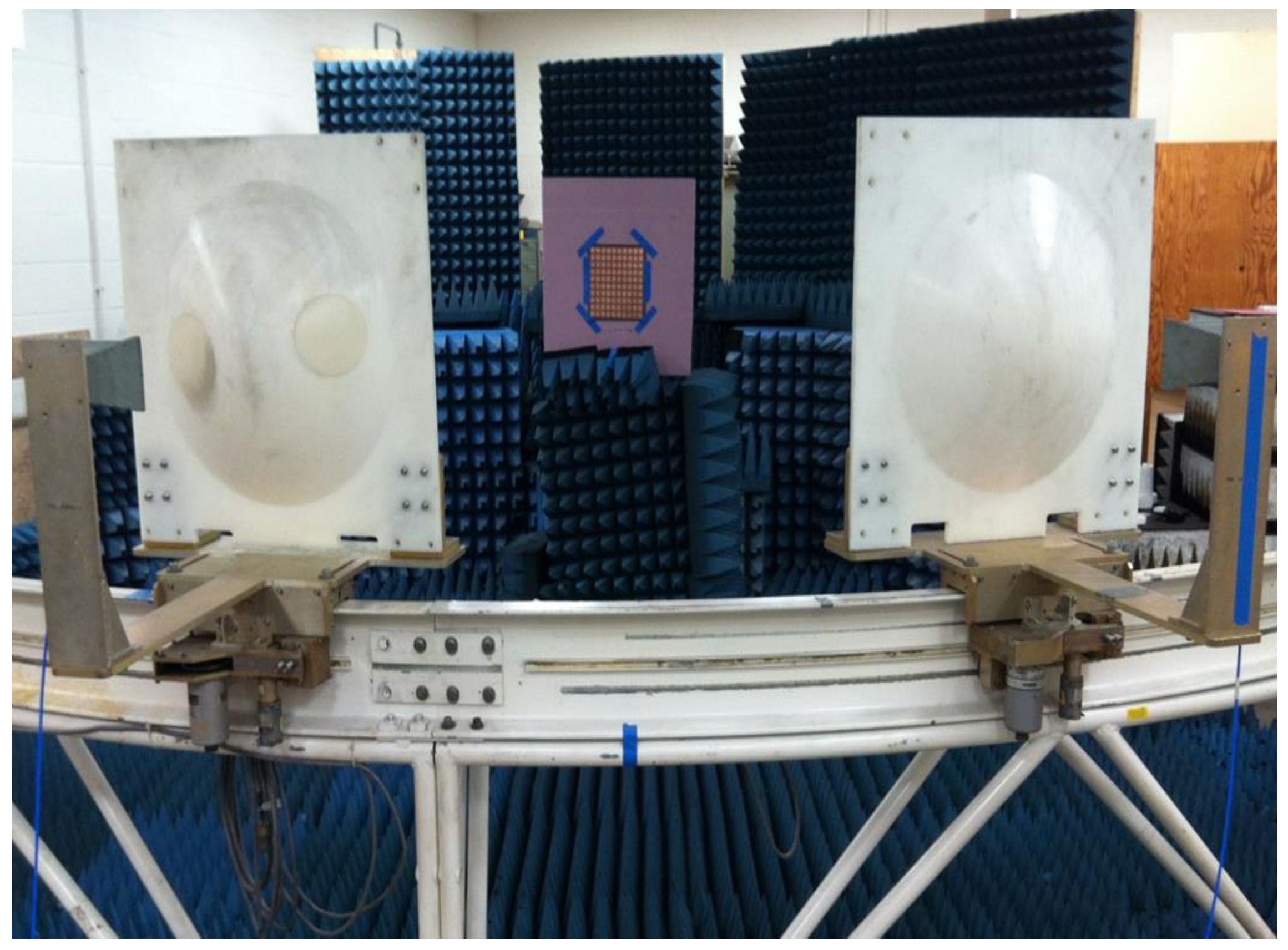}%
\label{archrangepic}}
\caption{(a) Close-up view of the reflectarray (b) Experimental setup for bistatic measurement with reflectarray at the center of the arch.}
\label{archrange}
\end{figure}

A 10x10 element reflectarray antenna was fabricated based on the proposed unit cell design. See Fig. \ref{archrange}. The dimensions and periodicity of the unit cell, described in section \MakeUppercase{\romannumeral 3} (A), correspond to a square array of 220~mm~x~220~mm. The substrate and the biasing circuit used in the full array are the same as that used in the single unit cell prototype described in section \MakeUppercase{\romannumeral 3} C. The biasing network for the varactor was fabricated on an additional substrate and glued behind the ground plane of the array. With this prototype, instead of controlling each unit cell individually to achieve a 3-D steerable beam, only the 2-D beam steering ability is investigated. Therefore, each column of the reflectarray is biased with the same voltage. Hence, ten digital voltage regulators were used to accurately bias the array elements.

A bistatic measurement was performed using an arch range at Michigan State University \cite{Perry} in order to measure the radiation pattern. TEM-horn antennas were connected to an Agilent E5071c network analyzer to measure the transmission coefficients S$_{21}$. Dielectric lenses were placed in front of the horn antennas to generate a focused beam with uniform phase. At 3~GHz and 6~GHz, the diameters of the beam are approximately 43 cm and 31 cm respectively. Note that in the lower frequency band, the beam size is larger than the array, which may cause a reduction in the measured S$_{21}$. Moreover, diffraction at the edge of the antenna will also bring a further reduction in the total transmitted power.

The VNA was calibrated from 1.5~GHz to 7.5~GHz with 1601 frequency points. The calibration of the measurement system was then conducted by finding the system response. The response of the system was calculated by considering the canonical problem of scattering off a metal sphere. The response function was obtained by comparing the experimental results with the analytical solution (Mie theory) of the scattering problem. The details of the calibration procedure is explained in \cite{Perry}.
The antenna array was mounted on a sheet of Styrofoam and placed at the center of the arch range. The arch range has a radius of 3.53~m. The center of the array was adjusted to be at the same height as the horn antennas.  The array and the transmitter were kept stationary and the direction of the incident wave was adjusted to be perpendicular to the surface of the array. Fig. \ref{archrange} shows the experiment setup with the antenna in place. The full radiation pattern of the antenna can be measured by moving the receiving antenna (the antenna on the right in Fig. \ref{archrange}) along the arch rail. It is important to mention that, due to the large profile of the dielectric lens, the receiving and transmitting antennas cannot be co-located.  Therefore, a gap in the antenna pattern appears from approximately -15$^{\circ}$ to 15$^{\circ}$. However, this gap does not affect the measurement at 0$^{\circ}$ for the co-polarization component.

\begin{figure}[!ht]
\begin{center}
\includegraphics[keepaspectratio = true, width = 3.5in, clip = true]{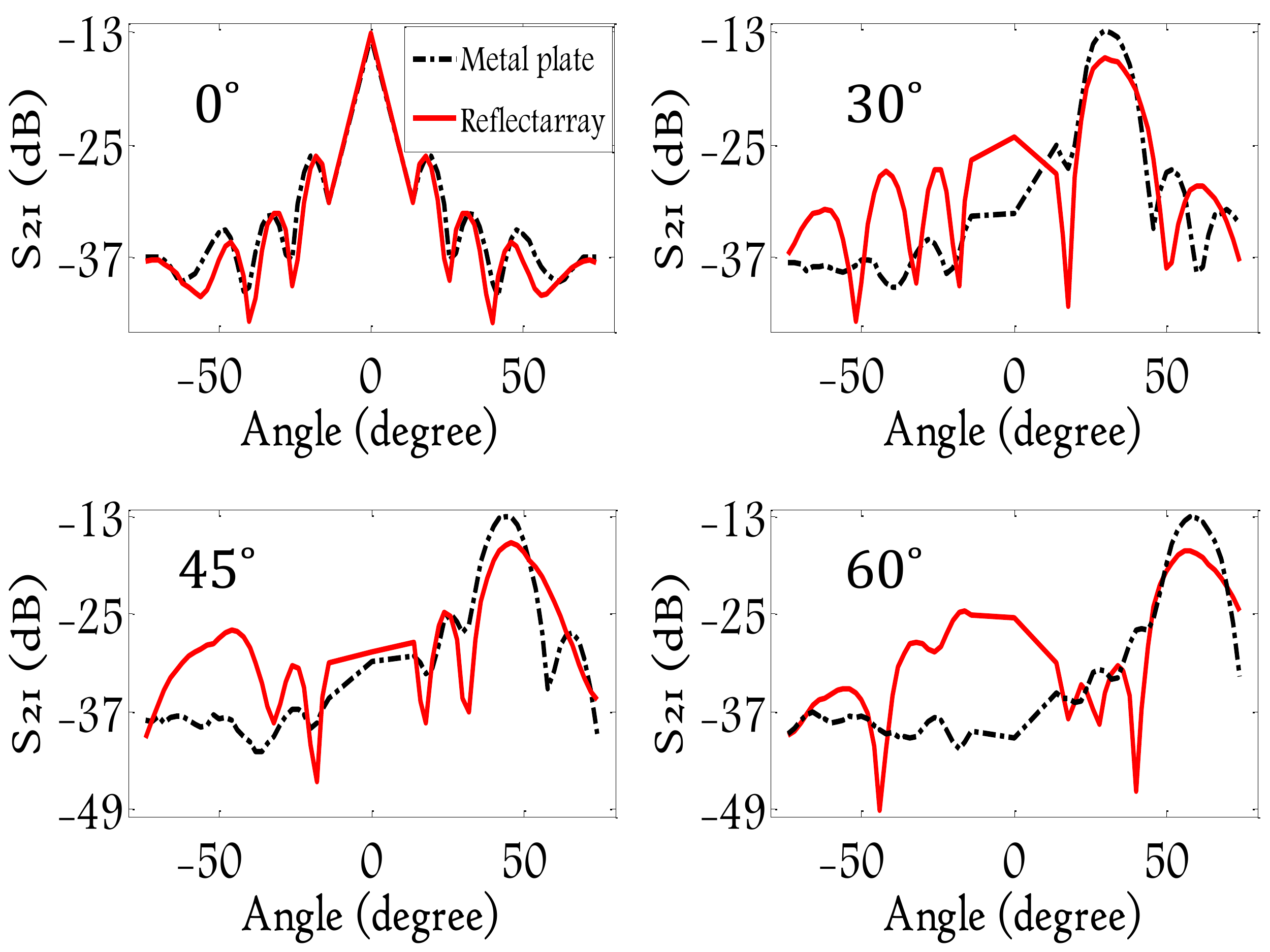}
\caption{Beam steering at 6.12~GHz.
}  \label{fbandstr}
\end{center}
\end{figure}

Different voltage configurations were applied to the reflectarray in order to steer the main beam in multiple directions. Fig. \ref{fbandstr} shows four different cases were the beam is steered onto the angles 0$^{\circ}$, 30$^{\circ}$, 45$^{\circ}$ and 60$^{\circ}$ at 6.12~GHz (in the higher frequency band). We define the reflection efficiency of the array as the ratio of the maximum of the transmission coefficient of the array to that of a metal plate. The metal plate the same dimensions as the reflectarray was manually rotated in order to steer the beam in the same angle as the array.

\begin{figure}[!ht]
\begin{center}
\includegraphics[keepaspectratio = true, width = 3.5in, clip = true]{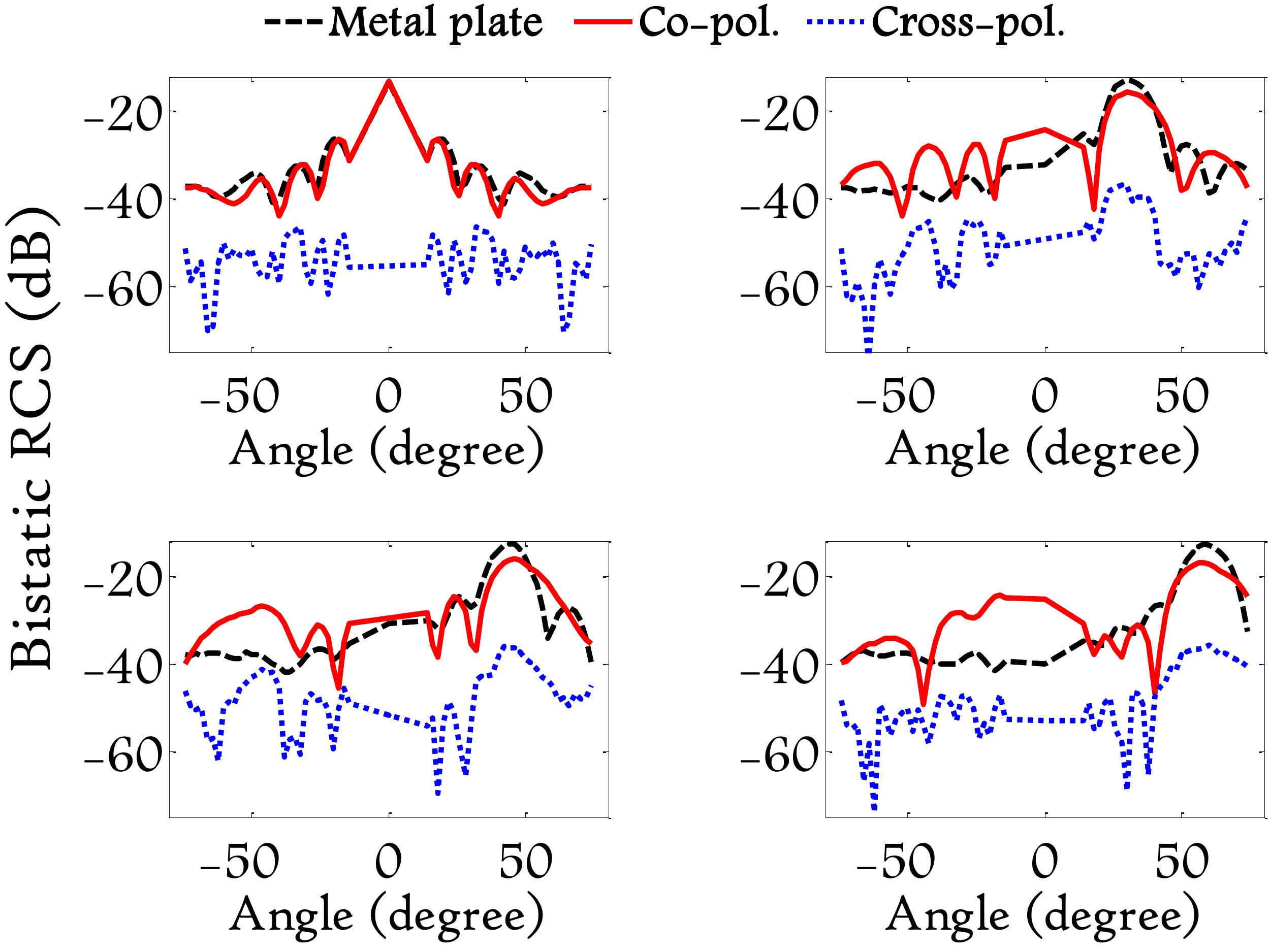}
\caption{Beam steering at 6.12~GHz with the cross polarized component.
}  \label{fbandstrCX}
\end{center}
\end{figure}

When the array is not biased, the measured pattern is similar as that of the metal plate, with the main beam reflected at 0$^{\circ}$. The triangular shape of the main beam is due to the described measurement gap of the system. At 30$^{\circ}$, reflection efficiency, the gain difference between the beam reflected by the array (red curve) and the one reflected by the metal plate (black curve) is -2.9~dB. At 45$^{\circ}$, the gain difference between the main lobes is -3.2~dB and at 60$^{\circ}$ is -4.2~dB. The increase of the power loss for beams steered at large angles is expected and is observed in \cite{Hum2}. This is because the overall return loss of the unit cells under the voltage combination for large steering angles is greater than that of smaller steering angles. In addition, It is obvious that the beam can be deflected into negative angles by simply reversing the applied voltage. This results in an effective beam tuning range of 120$^{\circ}$. The measured cross polarized component is shown in Fig. \ref{fbandstrCX}. The cross polarized component is expected to be low as explained in section \MakeUppercase{\romannumeral 3}. At the beam peak, the cross polarized component is at least 19 dB lower for all steering angles. It was observed that for the voltage configuration applied, within 6-6.2 GHz, the gain reduction is less than 1.5 dB for any steering angles below 60$^{\circ}$. This behavior indicates that the phase relationships are maintained in a narrow-band interval. Therefore, a new voltage configuration should be applied to steer the beam in a different frequency range.

\begin{figure}[!ht]
\begin{center}
\includegraphics[keepaspectratio = true, width = 3.5in, clip = true]{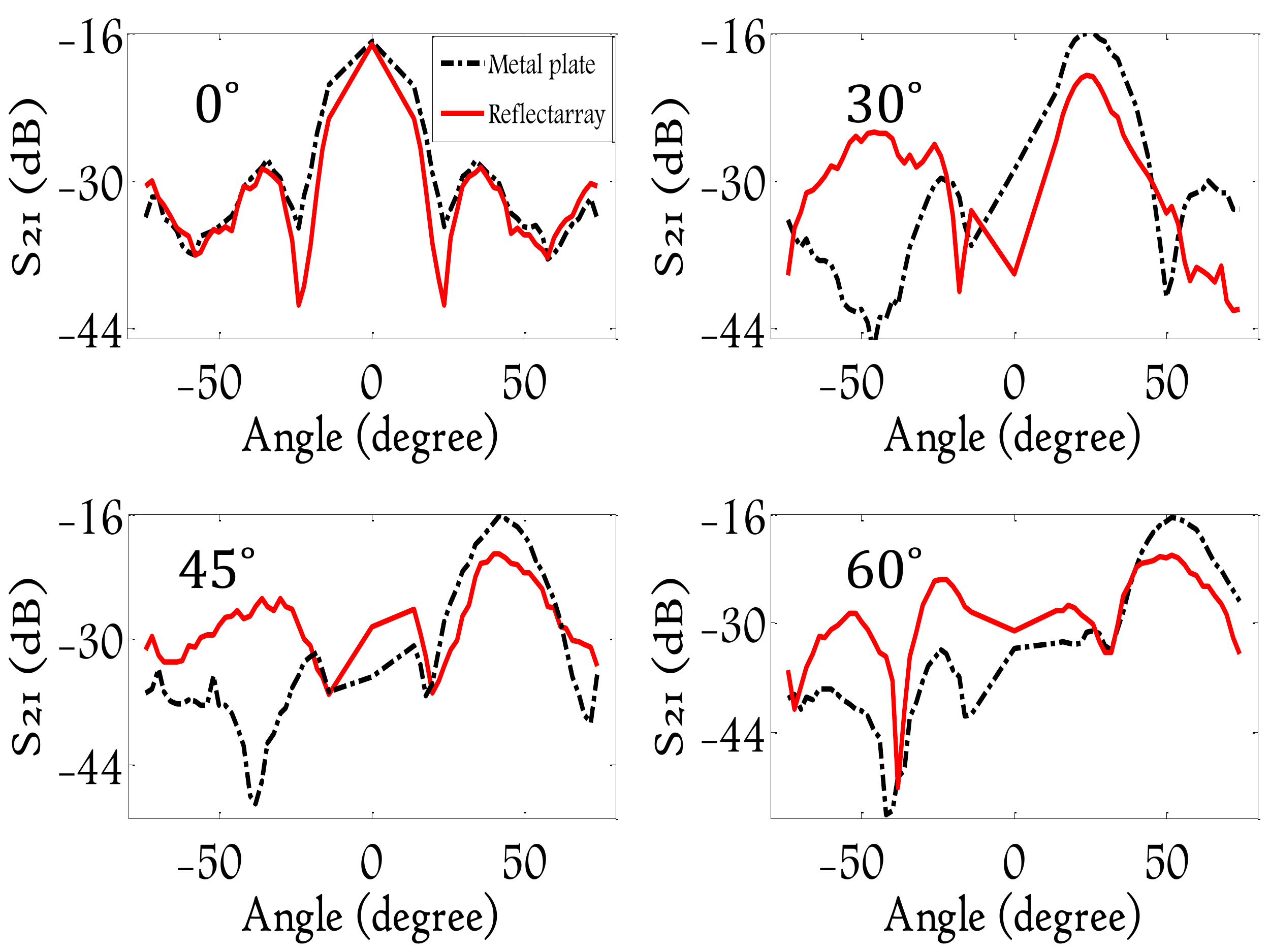}
\caption{Beam steering at 3.38~GHz.
}\label{sbandstr}
\end{center}
\end{figure}

Fig. \ref{sbandstr} shows beam steering at 3.38~GHz (in the lower frequency band). Since the unit cell is more lossy (see section \MakeUppercase{\romannumeral 3}) the efficiency of the array is in general lower in this frequency band. However, the beam could still be deflected up to almost 60$^{\circ}$.

\begin{figure}[!ht]
\begin{center}
\includegraphics[keepaspectratio = true, width = 3.5in, clip = true]{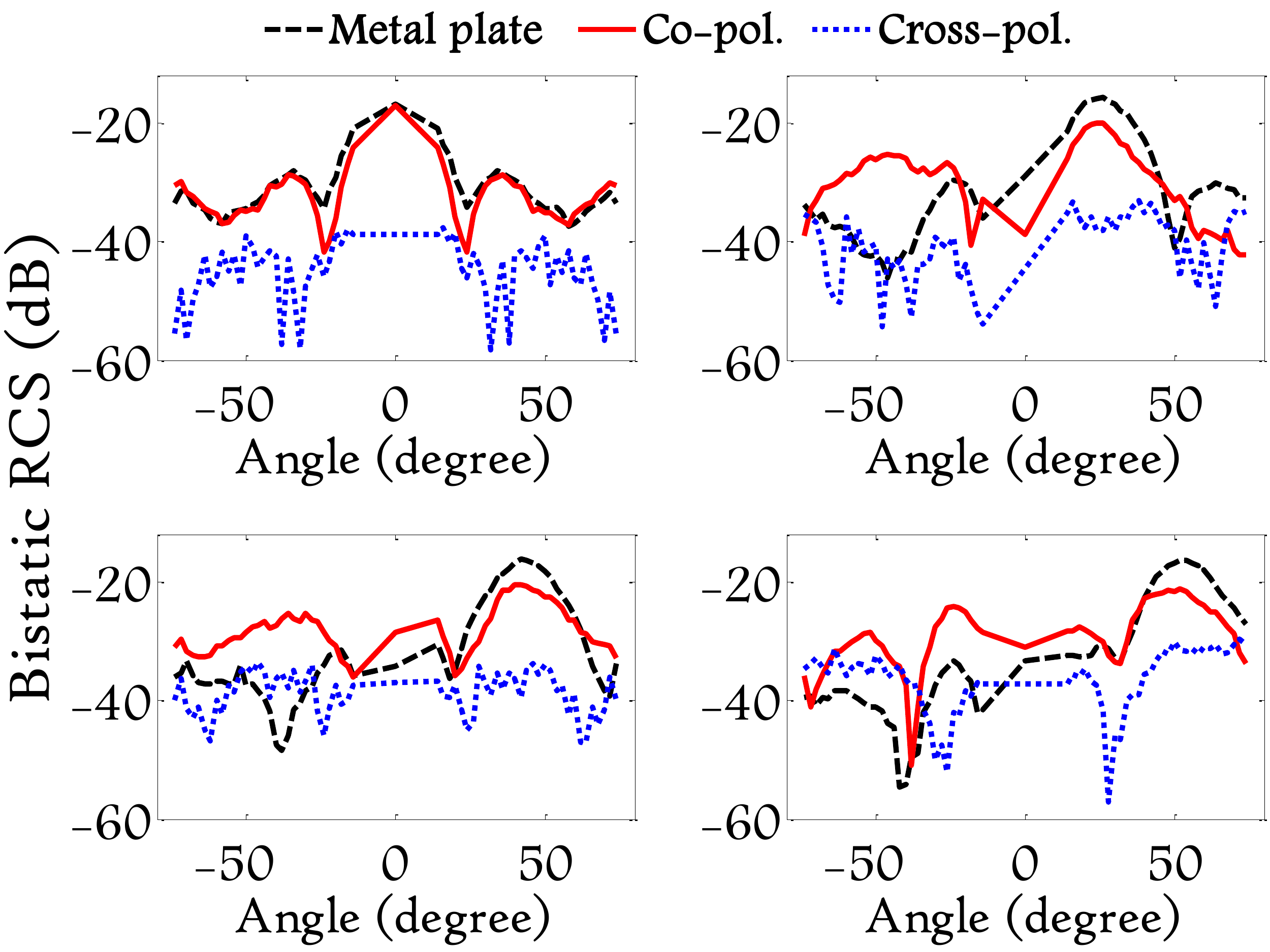}
\caption{Beam steering at 3.38~GHz with the cross polarized component.
}\label{sbandstrCX}
\end{center}
\end{figure}

As in the case of steering at higher frequency, when the array is turned off it behaves similarly to a metal plate. Note that the maximum of the main beam has been decreased comparing to the measurements at 6.12 GHz. This is due to the fact that the incident beam is more focused at higher frequencies and hence a larger portion of the power is reflected by both the array and the metal plate. The key element in evaluating the efficiency of the array is the relative difference between the metal plate and the array. The efficiency is -4.15 dB, -4.25~dB and -4.8~dB at 30$^{\circ}$, 45$^{\circ}$ and 60$^{\circ}$, respectively. To the best of our knowledge, such high beam steering in a dual-band tunable reflectarray has not yet been reported in literature. In general, a higher cross polarization component is observed in the S-band (see Fig. \ref{sbandstrCX}). A difference of at least 8 dB between the two polarizations is observed at the beam peak at all steering angles. Comparing to the F-band, the operating frequency range is relatively narrow. This is expected as it was seen in the measurement of the unit cell at S-band which indicated the narrow-band nature of such structure.

\section{Conclusion}

This paper presents the design of a unit cell with tuning ability over a wide range of phase. The unit cell operates at two frequency bands and requires only one varactor diode to dynamically alter the phase of the scattered field. This significantly reduces the manufacturing cost of the array compared to more complicated systems that require multiple varactors. The unit cell was evaluated using both full-wave simulations and equivalent circuit modeling. The equivalent circuit provides a simple description of the unit cell in terms of passive circuit elements, and requires far less computation time compared to the full wave simulations. The unit cell was evaluated experimentally by placing it into a waveguide and measuring the reflection coefficient. A maximum phase shift of 335$^{\circ}$ was achieved in the upper band, and 340$^{\circ}$ in the lower band.

The 10x10 reflectarray built using the proposed unit cell shows excellent steering capabilities. The beam can be deflected up to $\pm$60$^{\circ}$ in both frequency bands. This steering range has not yet been reported in literature. This unit cell design can be expanded to a larger scale to allow for a greater range of steering capabilities. Moreover, by individually biasing each unit cell, dynamic 3-D beam steering should be achievable. This is left for future study.

%








\end{document}